\documentclass[nofootinbib,prd,aps,onecolumn,preprintnumbers,amsmath,amssymb,superscriptaddress]{revtex4-2}
\usepackage{amsmath}
\usepackage{graphicx}
\usepackage{float}
\usepackage{dcolumn}
\usepackage{appendix}
\usepackage{bm}
\usepackage{amssymb}
\usepackage{latexsym}
\usepackage{subfigure} 
\usepackage{epstopdf}
\usepackage{threeparttable}
\usepackage{booktabs}
\usepackage{tabularx}
\usepackage{mathrsfs}
\usepackage[normalem]{ulem}        
\usepackage{cancel}
\usepackage{hhline}
\usepackage{multirow}
\usepackage{color}
\usepackage[colorlinks,linkcolor=magenta,anchorcolor=blue,citecolor=green]{hyperref}
\usepackage{url}
\usepackage{pifont}
\usepackage{makecell}
\def\be{\begin{equation}}
\def\ee{\end{equation}}
\def\bea{\begin{eqnarray}}
\def\eea{\end{eqnarray}}

\begin{document}
\title{Tensor Perturbations from Bounce Inflation Scenario in $f(Q)$ Gravity}

\author{Kun Hu}
\affiliation{Institute of Astrophysics, Central China Normal University, Wuhan 430079, China}
\affiliation{School of Physics, Huazhong University of Science and Technology, Wuhan, 430074, China}

\author{Tanmoy Paul}
\affiliation{National Institute of Technology Jamshedpur, Department of Physics, Jamshedpur - 831 014, India}

\author{Taotao Qiu}
\email{qiutt@hust.edu.cn (corresponding author)}
\affiliation{School of Physics, Huazhong University of Science and Technology, Wuhan, 430074, China}

\begin{abstract}
In this paper, we construct a bounce inflation cosmological scenario in the framework of the modified symmetric teleparallel gravity, namely $f(Q)$ theory, and investigate the tensor perturbations therein. As is well-known, the tensor perturbations generated in the very early Universe (inflation and pre-inflation regions) can account for the primordial gravitational waves (PGWs) that are to be detected by the next generation of GW experiments. We discuss the stability condition of the tensor perturbations in the bounce inflation process and investigate in detail the evolution of the perturbation variable. The general form of the tensor power spectrum is obtained both for large as well as small scale modes. As a result, we show both kinds of modes (short or long wavelength modes), the tensor spectrum may get a positive tilt in the parametric range where the tensor perturbation proves to be stable --- this interestingly hints an enhancement of gravitational waves' amplitude in the background of the $f(Q)$ bounce-inflation scenario. Moreover, we study the LQC-like scenario as a specific case of our model, in which, the primordial tensor power spectrum turns out to be nearly scale-invariant on both small and large scales.
\text{PACS numbers: 04.20.Dw, 04.50.Kd, 98.80.-k}
\end{abstract}

\maketitle

\section{Introduction}
It is well-known that our Universe, described by the Standard Big Bang Theory, suffered from the notorious problem of Cosmic Singularity \cite{Borde:1993xh} at the time of its birth. More precisely, as has been proved by S. Hawking et al. in the 1970s, the Universe will inevitably meet the singularity if it is 1) described by 4D classical General Relativity, 2) satisfying the Null Energy Condition (NEC) \cite{Hawking:1970zqf,hawking2023large}. In the vicinity of the singularity, physical quantities such as energy density blow up, therefore failing to present the real nature. Nonetheless, one of the simple and feasible ways of avoiding such a singularity is to have the universe start with a contracting phase, and bounce to the expanding one as we observed \cite{Novello:2008ra,Battefeld:2014uga,Cai:2014bea}. Such a scenario is usually dubbed the ``cosmic bounce".

To realize a cosmic bounce, at least one of the two conditions mentioned above needs to be broken. It might be more straightforward to break the NEC by including some exotic matter such as ghost field \cite{Cai:2007qw}, but in order to avoid releasing the ghost instability, the models should be delicately designed. There are examples made of ghost condensate mechanism \cite{Lin:2010pf}, Horndeski theories \cite{Qiu:2011cy, Easson:2011zy, Cai:2012va, Qiu:2013eoa, Qiu:2015nha}, DHOST theories \cite{Ilyas:2020qja,Zhu:2021whu,Ilyas:2020zcb} or physical investigation of the space-time defect \cite{Battista:2020lqv}. However, one can also obtain the bounce by modifying Einsein's General Relativity, namely referring to modified gravity theories, where the effective NEC is also broken. This includes non-minimal coupling bounce \cite{Qiu:2010ch, Qiu:2010vk}, $f(R)$ bounce \cite{Carloni:2005ii,Odintsov:2020zct}, $f(R,\mathcal{G})$ bounce \cite{Nojiri:2022xdo,Odintsov:2022unp}, $f(T)$ bounce \cite{Cai:2011tc, Hohmann:2017jao, Qiu:2018nle}, $f(Q)$ bounce \cite{Bajardi:2020fxh, Mandal:2021wer, Agrawal:2021rur, Shaikh:2022auv,Gadbail:2023loj,Agrawal:2022vdg}, Loop-quantum bounce \cite{Date:2004fj} and so on.   

Compared to other modified gravity theories, the bounce scenario in $f(Q)$ theory has been less studied, especially, the perturbation analysis of $f(Q)$ bounce scenario is far from sufficient. This motivates the work in the current paper. The $f(Q)$ theory is one of the interesting alternative theories of gravity, which extends from a metric theory to a metric-affine theory~\cite{BeltranJimenez:2017tkd}. It is well-known that in GR, the affine connection $\varGamma^\lambda_{~\mu\nu}$, and therefore the Ricci scalar $R$ of the gravity action, can be written in terms of metric only, making GR a pure metric theory. (In such case, $\varGamma$ is known the Christoffel symbol). This is because of two conditions: 1) torsionlessness, namely the connection is symmetric, $\varGamma^\lambda_{~\mu\nu}=\varGamma^\lambda_{~\nu\mu}$. 2) metric compatibility, namely the covariant derivative on metric vanishes, $\nabla_\lambda g_{\mu\nu}=0$. However, if one breaks one of the two conditions, the metric and connection become independent of each other, and the theory becomes metric-affine theory, in which the metric and the affine connection are both independent fundamental variables. To be precise, if one breaks condition 1) while keeping condition 2), the theory will become a theory with torsion, with asymmetric connection; In contrast, if one breaks condition 2) while maintaining condition 1), one will get another gravity theory with symmetric connection and no torsion but instead have non-metricity.

Nonetheless, both the two kinds of gravity theories can be made equivalent to GR, which has obtained great success at least in the range of low energy. For torsion theory, the action can be made of the torsion scalar $T$:
\begin{equation}
S=\int d^4x\, e T~,~T\equiv\frac{1}{4}T_{\lambda\mu\nu}T^{\lambda\mu\nu}+\frac{1}{2}T_{\lambda\mu\nu}T^{\mu\lambda\nu}-T_\lambda T^\lambda~,
\end{equation}
where we define the torsion tensor: $T^\lambda_{~\mu\nu}\equiv\varGamma^\lambda_{~\mu\nu}-\varGamma^\lambda_{~\nu\mu}$, which is to describe how the hypersurfaces twist about a vector after a parallel transportation, and $T_\mu\equiv T^\lambda_{~\mu\lambda}$, $e$ is the volume element given by the tetrad. This is dubbed as {\it Teleparrallel Equivalent of General Relativity} (TEGR) \cite{Aldrovandi:2013wha}. Similarly, for non-metricity theory, the action can be made of the nonmetricity scalar $Q$:
\begin{equation}
S=\int d^4x\,\sqrt{-g}Q~,~Q\equiv \frac{1}{2}Q_{\mu\nu\lambda}Q^{\lambda\mu\nu}-\frac{1}{4}Q_{\mu\nu\lambda}Q^{\mu\nu\lambda}+\frac{1}{4}Q_\lambda Q^\lambda-\frac{1}{2}Q_\lambda \tilde{Q}^\lambda~,~\label{Eq.nonmetricity scalar}
\end{equation}
where we define the non-metricity tensor: $Q_{\lambda\mu\nu}\equiv \nabla_\lambda g_{\mu\nu}$, which is to describe the variation of the norm of vector after parallel transportation, and its traces $Q_\lambda\equiv g^{\mu\nu}Q_{\lambda\mu\nu}$, $\tilde{Q}_\lambda\equiv g^{\mu\nu}Q_{\nu\mu\lambda}$. This is known to be {\it Symmetric Teleparrallel Equivalent of General Relativity} (STEGR) \cite{Nester:1998mp}.  It has been proved in refs.~\cite{Iosifidis:2018zjj,BeltranJimenez:2019esp} that these two actions, as well as the Einstein-Hilbert action of GR, only differ by a total derivative from each other, therefore they can share many of the physical properties and phenomenological predictions. However, if the actions are furtherly modified to be any functions of these scalars, this is no longer the case because the total derivatives inside the functions can no longer be thrown out. Therefore, the modified gravity theories $f(R)$, $f(T)$ and $f(Q)$ are completely different theories, thus may give rise to very different (and interesting) phenomenons.

There has already been a lot of discussion on $f(Q)$ theory and its various applications. For the development of the theory itself, we see ref.~\cite{Gakis:2019rdd} for the construction of a conformal invariant theory, and ref.~\cite{BeltranJimenez:2019odq,BeltranJimenez:2021auj} for a generalization to the so-called general teleparallel quadratic tensor theories. Also see refs.~\cite{Zhao:2021zab} for a discussion on the covariant formulation. For cosmological applications, refs.~\cite{Lu:2019hra, Xu:2020yeg, Hassan:2021egb, Mandal:2021bfu, Fu:2021rgu, Khyllep:2021pcu, Anagnostopoulos:2021ydo, Lin:2021uqa, Shiravand:2022ccb, Capozziello:2022tvv} investigated various cosmological evolutions, while refs.~\cite{BeltranJimenez:2019tme, Najera:2021afa} studied the cosmological perturbations. Especially, the gravitational waves generated by the theory are studied in refs.~\cite{Hohmann:2018wxu, Soudi:2018dhv, Conroy:2019ibo, Zhao:2019xmm, Chen:2022wtz}. Moreover, there are a lot of constraints on this theory from observations, see refs.~\cite{Lazkoz:2019sjl, Barros:2020bgg, Frusciante:2021sio, Arora:2021jik}. See also refs.~\cite{Golovnev:2018red, Heisenberg:2018vsk} for comprehensive reviews. 

Recently, in ref.~\cite{Hu:2022anq}, some of our authors studied the physical degrees of freedom of the perturbative $f(Q)$ theory in {\it coincident gauge}, by performing the Hamiltonian analysis approach. Our result shows that different from the usual GR theory, the $f(Q)$ theory in such a gauge can have many more additional degrees of freedom, which might be attributed to the violation of the general covariance. The additional degrees of freedom are basically scalar and vector ones, which are still under investigation. Nevertheless, it is much more evident that only two tensor degrees of freedom remain, since there is no room for redundant ones, which can be seen just from the Scalar-Vector-Tensor (SVT) decomposition. Therefore the tensor perturbations of this theory can be easily determined. Moreover, with the emergence of more and more Primordial Gravitational Waves (PGW) experiments such as NANOGrav \cite{NANOGrav:2020bcs}, AliCPT \cite{Li:2017lat, Li:2017drr}, CMB-S4 (\href{https://cmb-s4.org/}{https://cmb-s4.org/}) and LiteBIRD (\href{https://www.isas.jaxa. jp/missions/spacecraft/future/litebird.html}{https://www.isas.jaxa. jp/missions/spacecraft/future/litebird.html}), the investigation of tensor perturbations of the early universe models becomes more and more striking. Recently released data of Pulsar Timing Array from NANOGrav 15yr \cite{NANOGrav:2023gor}, EPTA \cite{Antoniadis:2023rey}, PPTA \cite{Reardon:2023gzh} and CPTA \cite{Xu:2023wog} shows the hint of stochastic GW background at 2 to $4.6\sigma$ level. Therefore, in the following, we will investigate the construction of a  bounce (inflation) scenario in the framework of $f(Q)$ theory, as well as the stability of the tensor perturbation and its evolution process through the bounce. 

The contents are arranged as follows. In Sec. \ref{sec2} we briefly review the basic formulation of $f(Q)$ theory. In Sec. \ref{sec3} we construct a model from $f(Q)$ theory with bouncing behavior at the background level. Sec. \ref{sec4} devotes itself to the perturbations generated in such a bounce model. Sec. \ref{sec5} comes to the conclusions and outlook. Throughout the paper, we work in units with $c =\hbar = 1$ and for simplicity, we set the gravitational constant $\kappa ^2=8\pi G_N=1$.

\section{the $f(Q)$ theory}
\label{sec2}
The most general action of $f(Q)$ gravity is given by the following expression~\cite{BeltranJimenez:2019tme,Lu:2019hra,Harko:2018gxr}
\begin{equation}
S=\int d^4 x\,\sqrt{-g} \left[\frac{1}{2} f(Q)+\mathcal{L}_m\right]~, \label{Eq.action}
\end{equation}
where $g$ is the determinant of the metric $g_{\mu\nu}$, $f(Q)$ is an arbitrary function of the nonmetricity scalar $Q$ appeared in~\eqref{Eq.nonmetricity scalar} and $\mathcal{L}_m$ stands for the matter lagrangian density.
We consider the general linear affine connection as
\begin{equation}
	\varGamma^{\alpha}_{\,\mu \nu}= \left\{ {}^{\, \alpha}_{\mu \nu} \right\} + K_{\,\mu \nu}^{\alpha} + L_{\,\mu \nu}^{\alpha}~.\label{Eq.connection}
\end{equation}
Here, the first term of the right-hand side of the above equation is the well-known Levi-Civita connection
\begin{equation}
	\left\{ {}^{\, \alpha}_{\mu \nu} \right\}
	=\frac{1}{2} g^{\alpha \lambda}\left(\partial_\mu g_{\lambda \nu} + \partial_{\nu} g_{\mu \lambda} - \partial_{\lambda}g_{\mu \nu} \right)~,~\label{Levi-Civita connection}
\end{equation}
the second term is the contortion tensor $K_{\,\mu \nu}^{\alpha}$, which is a function of the torsion tensor $T^{\lambda}_{\,\mu \nu}$
\begin{equation}
	K^{\alpha}_{\,\mu \nu} 
	= \frac{1}{2} g^{\alpha \lambda} \left( T_{\lambda \mu \nu} + T_{\mu \nu \lambda} - T_{\nu \lambda \mu} \right)~,
\end{equation}
and the last term $ L_{\mu \nu}^{\alpha} $ is disformation tensor, and it is composed of nonmetricity tensor $Q_{\lambda\mu\nu}$
\begin{align}
	L^{\alpha}_{\, \mu \nu} 
	= -\frac{1}{2} g^{\alpha \lambda} \left( Q_{\mu \lambda \nu} + Q_{\nu \lambda \mu} - Q_{\lambda \mu \nu} \right)~.
\end{align}
The nonmetricity tensor is defined by
\begin{equation}
Q_{\alpha \mu \nu} \equiv  \hat\nabla_{\alpha} g_{\mu \nu}=\frac{\partial g_{\mu \nu}}{\partial x^{\alpha}}-g_{\nu \sigma} {\varGamma}_{\mu \alpha}^{\sigma}-g_{\sigma \mu} {\varGamma}_{\nu \alpha}^{\sigma}~,\label{Eq: nonmetricity}
\end{equation}
where $\hat\nabla_{\alpha}$ denotes the covariant derivative associated with the general affine connection. Meanwhile, the space-time curvature is described by the usual Riemann tensor
\begin{equation}
	R_{\beta \mu \nu}^{\alpha}(\varGamma)=\partial_{\mu} \varGamma_{\nu \beta}^{\alpha}-\partial_{\nu} \varGamma_{\mu \beta}^{\alpha}+\varGamma^{\alpha}_{\mu \lambda} \varGamma_{\nu \beta}^{\lambda}-\varGamma_{\nu \lambda}^{\alpha} \varGamma_{\mu \beta}^{\lambda}~.\label{Eq.Ricci}
\end{equation}

Symmetric teleparallelism, by definition, requires free curvature and torsion, namely $R_{\beta \mu \nu}^{\alpha}(\varGamma)=0$ and $T_{\mu \nu}^{\alpha}=K_{\mu \nu}^{\alpha}=0$ respectively. Thereby, the form of general affine connection has been restricted as following
   \begin{equation}
	\varGamma_{\,\mu \nu}^\alpha=\frac{\partial x^\alpha}{\partial \xi^\lambda} \partial_\mu \partial_\nu \xi^{\lambda}~.\label{Eq:connection1}
\end{equation}
Equation~\eqref{Eq:connection1} turns out to be a pure-gauge connection and hence it is always possible to make it vanish by utilizing general coordinate transformation to set $\xi^\lambda=x^\lambda$. This gauge choice was named the coincident gauge.

Substituting Eq.~\eqref{Eq.connection} (with $K_{\mu \nu}^{\alpha}=0$) into Eq.~\eqref{Eq.Ricci} one can find that the Ricci scalar of the Einstein-Hilbert action $\mathring{R}$, composed of the Levi-Civita connection~\eqref{Levi-Civita connection}, can be expressed by the nonmetricity scalar plus a total divergence of two independent traces of the nonmetricity tensor, which become the boundary term after integration
\begin{equation}
	\mathring{R}=Q-{\nabla}_{\alpha}\left(Q^{\alpha}-\tilde{Q}^{\alpha}\right)~.\footnote{The notation ``$\circ$'' here denote the quantities related to General Relativity, and $\mathring{G}_{\mu \nu}=\mathring{R}_{\mu \nu}-\frac{1}{2}g_{\mu \nu}\mathring{R}$.}~\label{relation.Q.R}
\end{equation}

It is also useful to introduce the so-called nonmetricity conjugate or superpotential tensor 
\begin{equation}
P^\alpha{}_{\mu \nu}\equiv \frac{1}{2} \frac{\partial Q}{\partial Q_\alpha{ }^{\mu \nu}}=-\frac{1}{4} Q^\alpha{ }_{\mu \nu}+\frac{1}{2} Q_{\left(\mu \nu\right)}^\alpha+\frac{1}{4} g_{\mu \nu} Q^\alpha-\frac{1}{4}\left(g_{\mu \nu} \tilde{Q}^\alpha+\delta^\alpha_{(\mu} Q_{\nu)}\right)~,
\end{equation}
and the equation of motion can directly be obtained by taking the variation of the action towards metric $g_{\mu\nu}$, namely
\begin{equation}
\frac{2}{\sqrt{-g}} \nabla_{\alpha}\left(\sqrt{-g} f_{Q} P_{\ \mu \nu}^{\alpha}\right)+\frac{1}{2} g_{\mu \nu} f+f_{Q}\left(P_{\mu \alpha \beta} Q_{\nu}^{\ \alpha \beta}-2 Q_{\alpha \beta \mu} P_{\ \ \ \nu}^{\alpha \beta}\right)=T_{\mu \nu}~,\label{Eq.field equation}
\end{equation}
where the energy-momentum tensor of the matter is defined by
\begin{equation}
T_{\mu \nu}=-\frac{2}{\sqrt{-g}} \frac{\delta \sqrt{-g} \mathcal{L}_m}{\delta g^{\mu \nu}}~.
\end{equation}
Here and thereafter, $f_{Q}$ stands for the derivative of a function $f(Q)$ with respect to nonmetricity scalar $Q$. 
In particular, with the help of $P^{\alpha \lambda}{}_{ \lambda}=( Q^{\alpha}-\tilde{Q}^{\alpha} )/2 $, the relation~\eqref{relation.Q.R} and the identity
\begin{equation}
\nabla _{\alpha}\sqrt{-g}=\left( 2\sqrt{-g} \right) ^{-1}\frac{\partial g}{\partial g_{\mu \nu}}\nabla _{\alpha}g_{\mu \nu}=\frac{\sqrt{-g}}{2}Q_{\alpha}~,
\end{equation}
we can rewrite the Einstein tensor in GR in terms of superpotential and nonmetricity tensor~\cite{Bahamonde:2022esv},
\begin{equation}
\stackrel{\circ}{G}_{\mu \nu}=2 \nabla_\lambda P^\lambda{ }_{\mu \nu}-\frac{1}{2} Q g_{\mu \nu}+P_{\rho \mu \nu} Q^{\rho \sigma}{ }_\sigma+P_{\nu \rho \sigma} Q_\mu{ }^{\rho \sigma}-2 P_{\rho \sigma \mu} Q^{\rho \sigma}{ }_\nu~,~\label{Eq.Einstein tensor}
\end{equation}
Plugging the relation~\eqref{Eq.Einstein tensor} and into Eq.~\eqref{Eq.field equation}, the field equation can be written into a more compact form~\cite{Zhao:2021zab}: 
\begin{equation}
2 f_{QQ} P_{\mu \nu}^\alpha \partial_\alpha Q+\frac{1}{2} g_{\mu \nu}\left(f-f_{Q} Q\right)+f_{Q} \stackrel{\circ}{G}_{\mu \nu}= T_{\mu \nu}~.~\label{Eq.field equation2}
\end{equation}

In the case of STEGR, i.e., $f(Q)=Q$, the above equation reduces to Einstein’s equation in GR, thus the degrees of freedom from the affine connection do not enter into the field equation. But in a more general case, the $f(Q)$ is not the linear function of nonmetricity scalar, which gives a totally different evolution of the physical system from GR.

\section{Bounce Model within $f(Q)$ gravity}
\label{sec3}

\subsection{Background equations}
\label{subsecA}
In bouncing cosmology, the Universe begins from an epoch in the contracting phase in the far past, when it contracts to a very small but non-zero scale, and enters the expansion phase through a bounce, bridging with the thermal history of the expansion of the Universe that we observe today. Throughout the process, the scale factor of the Universe never reaches zero, and thus provides a possible way to avoid the singularity problem of the Universe. 

To begin with, we would have a brief analysis of the cosmological background evolution, considering the flat Friedmann-Lemaitre-Robertson-Walker (FLRW) background with the metric
\begin{equation}
ds^2=-dt^2+a^2\left( t \right) \delta _{ij}dx^idx^j~,~\label{Eq.FLRW}
\end{equation}
where $a(t)$ is the scale factor. On the other hand, if we regard the background of the Universe as a perfect fluid, then the stress-energy tensor takes the form of
\begin{equation}
    T_{\,\,\nu}^{(m)\mu}=\left( \rho_m +p_m \right) u^{\mu}u_{\nu}+p_m\,\delta _{\,\,\nu}^{\mu}~,
\end{equation}
where $\rho_m$ and $p_m$ are the energy density and pressure of the fluid, while $u^{\mu}=(1,0,0,0)$ is the fluid's four-velocity. More explicitly, we can write stress-energy tensor into a matrix form
\begin{equation}
	T^{(m)}_{\mu\nu}=\left[\begin{array}{cccc}
		\rho_m & 0 & 0 & 0 \\
		0 & a^{2}(t)p_m & 0 & 0  \\
		0 & 0 & a^{2}(t)p_m & 0  \\
		0 & 0 & 0 & a^{2}(t)p_m
	\end{array}\right]~.
\end{equation}

In the contracting epoch, the ``velocity" of the scale factor $\dot a(t)$ is negative, which causes the Hubble parameter $H\equiv \dot{a}/{a}<0$. However, it turns positive in the expanding phase. That means we must have $H=0$, $\dot{H}>0$ for the cosmological turnaround, which is exactly the bouncing point. In the case of Einstein's gravity with some matter fields, we have
\begin{equation}
\frac{\dot{H}}{H^2}=-\frac{3}{2}\left( \omega_m +1 \right)~,~\label{Eq.Fried}
\end{equation}
where $\omega_m\equiv p_m/\rho_m$ is the equation of state (EoS) of the matter field under consideration, it shows that only with $\omega_m=-\infty $ it has the power to produce the bounce, which clearly violates all the energy conditions. This clearly reflects that in the case of GR, the Universe can not achieve bounce without the presence of some extremely exotic matter. However, it is not the case in the framework of $f(Q)$ gravity, since the geometric effects of space-time can be interpreted as effective matter that violates the NEC.

For the $f(Q)$ gravity, we take the action form as in Eq.~\eqref{Eq.action} and work in the coincident gauge $\varGamma_{\ \mu \nu}^{\lambda} \equiv 0$, in which we have $Q_{\alpha \mu \nu}=\partial_{\alpha} g_{\mu \nu}$. Using metric~\eqref{Eq.FLRW} the equations of motion \eqref{Eq.field equation2} becomes
\begin{equation}
\begin{aligned}
6 f_Q H^2-\frac{1}{2} f & =\rho_m, \\
\left(12 H^2 f_{Q Q}+f_Q\right) \dot{H} & =-\frac{1}{2}(\rho_m+p_m)~.~\label{Eq.Friedmann equation}
\end{aligned}
\end{equation}
Comparing Eq.~\eqref{Eq.Friedmann equation} to the Friedmann equation in GR, we can rewrite the effective Friedmann equations as $ 3H^{2}=\rho_{\text{eff}} $ and $2\dot{H}+3H^2=-p_{\text{eff}}$, where $\rho _{\text{eff}}$ and $p_{\text{eff}}$ are effective density and pressure of the total fluid: 
\begin{equation}
\begin{aligned}
\rho _{\text{eff}}&=\frac{1}{2f_Q}\left( \rho_m +\frac{1}{2}f \right)~,\\
p_{\text{eff}}&=\frac{1}{f_Q}\left[ 12H^2f_{QQ}\dot{H}-\frac{f}{4}+\frac{p_m}{2}-\dot{H}f_Q \right] ~.
\end{aligned}
\end{equation}
Moreover, the EoS parameter of matter is $\omega_{m}=p_m/\rho_m$, which satisfies the continuity equation 
\begin{equation}
\overset{\cdot}{\rho_m}+3H\rho_m \left( 1+\omega_{m}  \right)=0~.~\label{continuity equation}
\end{equation}

It is also worth mentioning that, the relation between the nonmetricity scalar and Hubble parameter is simply
\begin{equation}
Q=6H^2(t)~,~\label{Eq.6H^2}
\end{equation}
which can be explicitly deduced from the definition of nonmetricity scalar~\eqref{Eq.nonmetricity scalar} for the background metric~\eqref{Eq.FLRW}. Therefore, one can make use of the Friedmann equations \eqref{Eq.Friedmann equation} to solve $f(Q)$ in terms of $H$. Define $f_H\equiv f_Q dQ/dH$ to replace $f_Q$ in the first equation in \eqref{Eq.Friedmann equation}, one gets
\begin{equation} 
f_{H}=\frac{1}{H}f(H)+\frac{2\rho_m(H)}{H}~,
\end{equation} 
the general solution of which is given by,
\begin{equation}
	f(H)=H\left[\sqrt{6}A + 2\int  \frac{\rho_m(H)}{H^2}\text{d} H\right] ~.~\label{Eq.differential equation}
\end{equation}
Here $A$ is an integration constant. However, in order to get the exact expression of $f$ in terms of $Q$ or $H$, the dependence of $\rho_m$ on $H$ must be known as well.

\subsection{Reconstructing the bounce action}
\label{subsecB}

\begin{figure*}[htb]
\centering
\includegraphics[height=6cm,width=8.0cm]{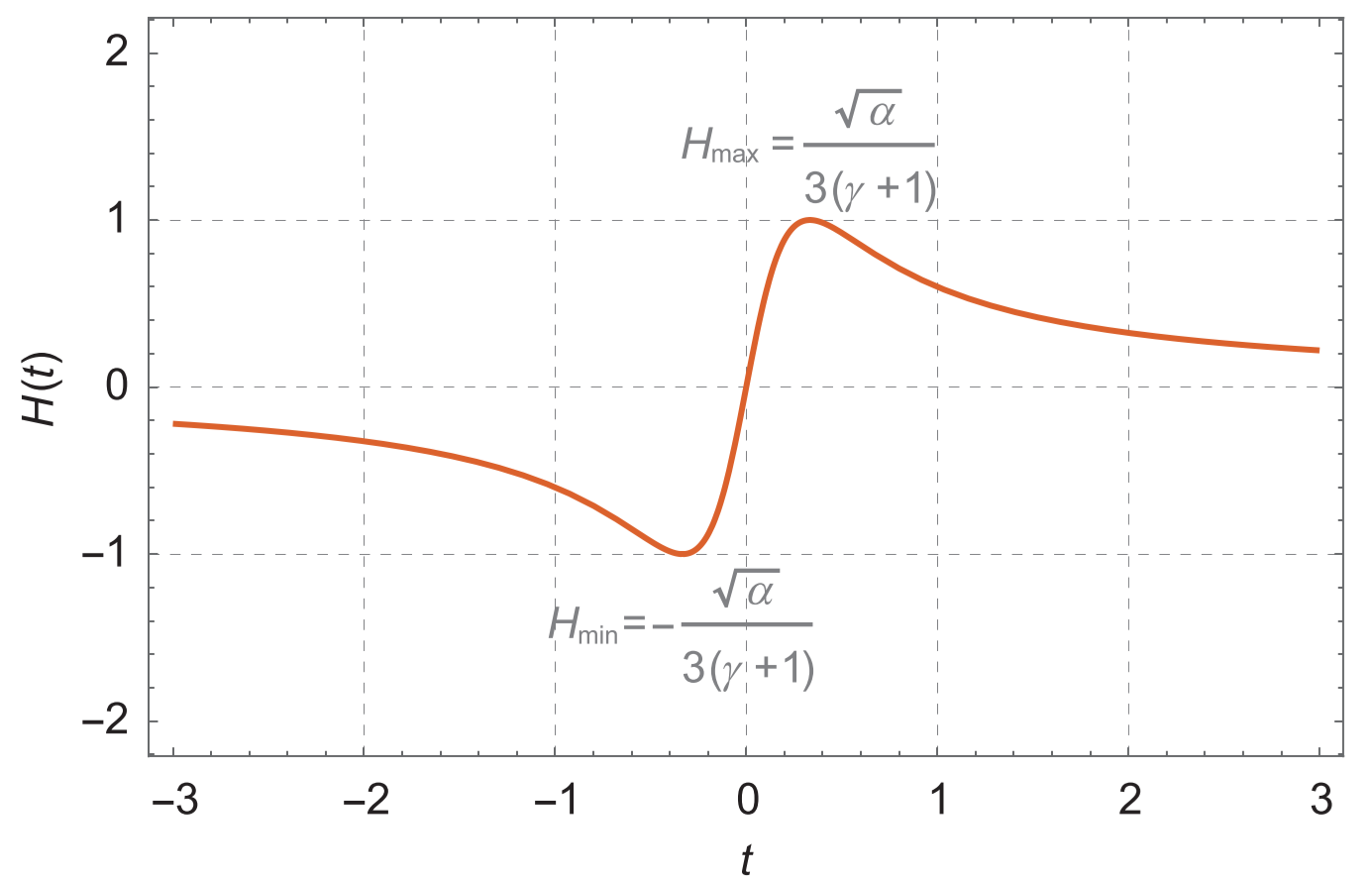}
\includegraphics[height=6cm,width=8.0cm]{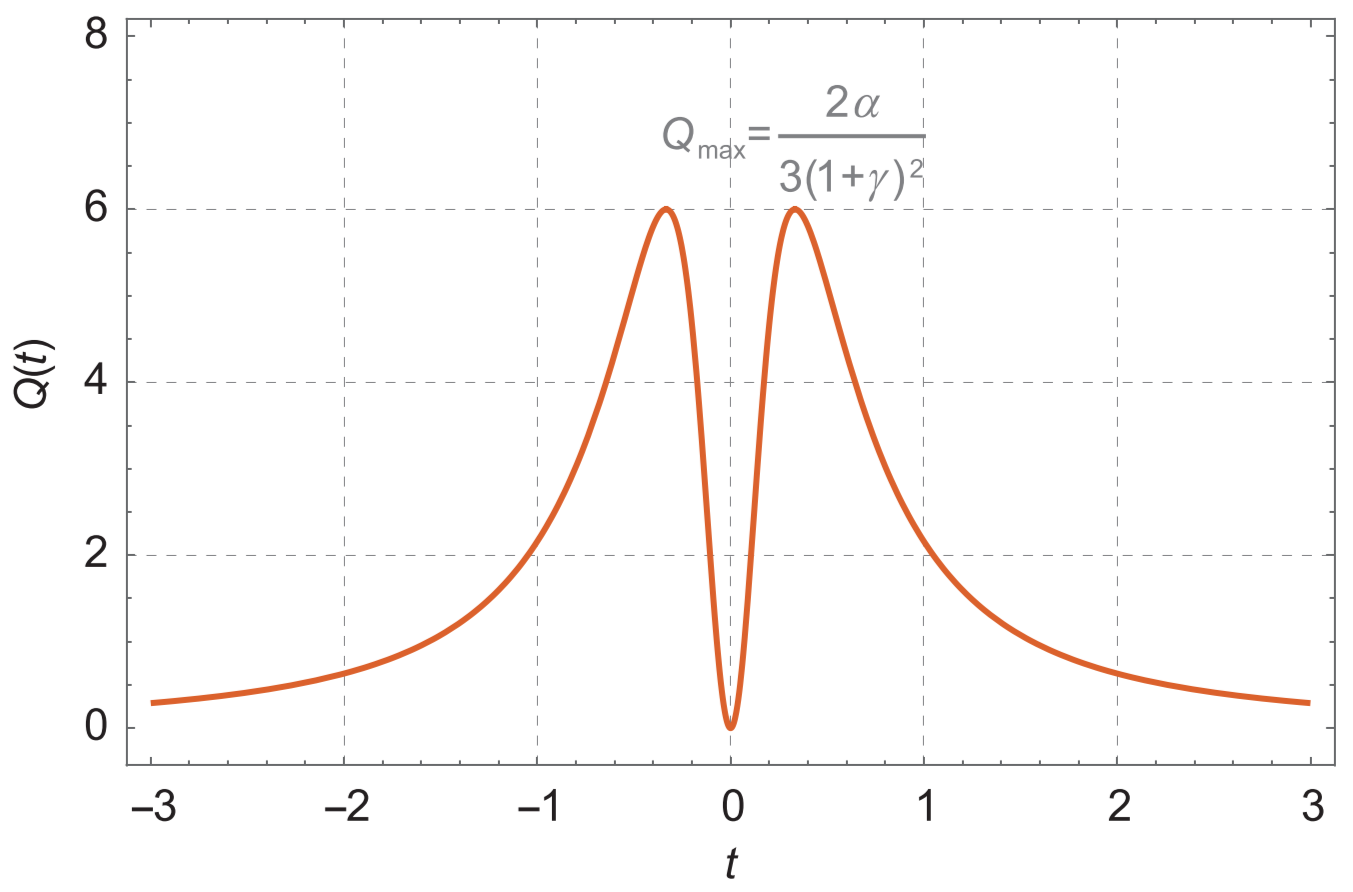}
\caption{The evolution plots of the Hubble parameter $H$  (left panel) and nonmetricity scalar $Q$ (right panel) in terms of the cosmic time $t$ for the scale fator~\eqref{Eq.scale factor} (the parameters are chosen as $\alpha=9$ and $\gamma=0$ for both plots). The Hubble parameter and nonmetricity scalar are both zero at the bounce point and meet their critical value $ H_{\max}/H_{\min} $ and $ Q_{\max} $ at $t=\pm 1/\sqrt{\alpha}$ respectively. When $t\rightarrow\infty$ the Hubble parameter and nonmetricity scalar both asymptotically go to zero.} 
\label{Q(t) and H(t)}
\end{figure*}

In order to realize a bounce scenario in this $f(Q)$ gravity, we apply the reconstruction approach and start with a specific, simple but quite natural scale factor with the form
\begin{equation}
  a\left(t\right)=a_{B}\left(1+\alpha \,t^{2}\right)^{\frac{1}{3(\gamma +1)}}~,~\label{Eq.scale factor}
\end{equation}
where $a_{B}$ is the scale factor at the bouncing point, $\alpha$ is a positive parameter that describes how fast the bounce takes place, and $\gamma$ is a constant parameter. The corresponding Hubble parameter is given by
\begin{equation}
	H(t)=\frac{\overset{\cdot}{a}}{a}=\frac{2\alpha t}{3\left( 1+\gamma \right) \left( 1+\alpha t^2 \right)}~.~\label{Eq.H(t)}
\end{equation}
We plot the Hubble parameter as well as the nonmetricity scalar in terms of $t$ in Fig. \ref{Q(t) and H(t)}.

Now, we can define the effective equation of state (EoS) as $\omega_{\text{eff}}=p_{\text{eff}}/\rho_{\text{eff}}$, so that Eq.~\eqref{Eq.Fried} still holds with $\omega_m\rightarrow \omega_{eff}$. Plugging the expression of Hubble parameter~\eqref{Eq.H(t)} into Eq.~\eqref{Eq.Fried}, we have
\begin{equation}
	\omega_{\text{eff}}\left( t \right) =\left( 1+\gamma \right) \left( 1-\frac{1}{\alpha  t^2} \right) -1~.
\end{equation}
We immediately see that for the late time Universe, i.e., $ |t|\gg 1/\sqrt{\alpha}$, we have the relation $\omega_{\text{eff}}\approx \gamma $, while when the Universe is near the bounce point, i.e., $ t\rightarrow 0$, $\omega_{\text{eff}}$ goes to $-\infty$.

With the help of parametrization form ~\eqref{Eq.H(t)}, one can firstly express the inverse relation of cosmic time $t$ in terms of $H$ as:
\begin{equation}
\begin{aligned}
	t_{1} &= \frac{\alpha-\sqrt{\alpha^{2}-9 \alpha H^{2}-18 \alpha H^{2} \gamma-9 \alpha H^{2} \gamma^{2}}}{3(\alpha H+\alpha H \gamma)}~,\\
	t_{2} &= \frac{\alpha+\sqrt{\alpha^{2}-9 \alpha H^{2}-18 \alpha H^{2} \gamma-9 \alpha H^{2} \gamma^{2}}}{3(\alpha H+\alpha H \gamma)}~.\label{Eq.t(H)}
 \end{aligned}
\end{equation}
The solutions have been separated into two branches. In the first branch $t_1$ increases (or decreases) with increasing (or decreasing) Hubble parameter, while the scenario is reversed in the second branch. From the Fig.~\ref{Q(t) and H(t)} one can see that, for $-1\leqslant \sqrt{\alpha}\, t\leqslant1 $, the relation $t=t(H)$ is given by the first branch; while for $-\infty<\sqrt{\alpha}\, t \leqslant-1 $ and $ 1\leqslant \sqrt{\alpha}\, t< \infty$ the relation $t=t(H)$ is given by the second branch.

Note that the straightforward solution of energy conservation equation~\eqref{continuity equation} gives the evolution of matter 
\begin{equation}
\rho_m(H)=\rho_{B}\left(\frac{a_{B}}{a(t)}\right)^{3(\omega_{m}+1)}~,~\label{Eq.evolution of matters}
\end{equation}
with $\rho_{B}$ being its value at the bouncing point. Plugging the scale factor~\eqref{Eq.scale factor} and the expression~\eqref{Eq.t(H)} into Eq.~\eqref{Eq.evolution of matters}, for the time branches $t_1$ and $t_2$, the corresponding energy densities evolve as
\begin{equation}
\begin{aligned}
\rho_{m1} \left( H \right) &=\rho_{B}\left( \frac{Q_m-Q_m\sqrt{1-6H^2/Q_m}}{3H^2} \right)^n~, \\
\rho_{m2} \left( H \right) &=\rho_{B}\left( \frac{Q_m+Q_m\sqrt{1-6H^2/Q_m}}{3H^2} \right)^n~.~\label{Eq.density evolve}
\end{aligned}
\end{equation}
Here, for simplicity, we introduce factors $n\equiv -(1+\omega_{m})/(1+\gamma)$, and $Q_m\equiv 2\alpha/[3\left( 1+\gamma \right) ^2]$  which denotes the maximum value of $Q$ (see Fig. \ref{Q(t) and H(t)}).

The interesting fact is that if we choose $\gamma=\omega_m=0$ and $\rho _B=4\alpha/3$, the evolution Eq.~\eqref{Eq.density evolve} takes the form
\begin{equation}
	\rho_m \left( H \right) =\frac{6H^2}{1\pm \sqrt{1-9H^2/\alpha}}~,
\end{equation}
which has the same form as the effective Friedmann equation in the context of Loop Quantum Cosmology ($\text{LQC}$)~\cite{Ashtekar:2011ni}
\begin{equation}
3H^2=\rho_m \left( 1-\frac{\rho_m}{\rho _B} \right)~. ~\label{Eq.QLC}
\end{equation}
For $\alpha \rightarrow \infty$, it leads to $\rho _B\rightarrow\infty$ and thus Eq.~\eqref{Eq.QLC} reduces to the Friedmann equation in GR. The above analysis illustrates that our model reduces to $\text{LQC}$ when we take the special case $n = -1$. For the sake of generality, we keep the parameter $n$ arbitrary in the subsequent calculation. 

Inserting Eq.~\eqref{Eq.density evolve} and Eq.~\eqref{Eq.H(t)} into expression~\eqref{Eq.differential equation}, we can get the form of $f=f(H)$ in the present context. Finally, utilizing the relation~\eqref{Eq.6H^2} again, the reconstructed Lagrangian of $f(Q)$ is obtained:

For $-1\leqslant \sqrt{\alpha}\, t\leqslant1 $:
\begin{equation}
\begin{aligned}
f\left( Q \right) =A^{(near)}\sqrt{Q}-\rho_{B}2^{n+1}\left( 1+\sqrt{1-\hat{Q}} \right)^{-n}&+\sqrt{2}\rho_{B}n\hat{Q}\left( 1+\sqrt{1-\hat{Q}} \right) ^{-\frac{1}{2}}\\
 &\times {}_2\text{F}_1\left[ \frac{1}{2},\frac{3}{2}+n,\frac{3}{2},\frac{1}{2}-\frac{1}{2}\sqrt{1-\hat{Q}} \,\right]~,~\label{Eq.late time f(Q)}
 \end{aligned}
\end{equation}

For $-\infty<\sqrt{\alpha}\, t \leqslant-1 $ and $ 1\leqslant \sqrt{\alpha}\, t< \infty$:
\begin{equation}
\begin{aligned}
	f\left( Q \right) =A^{(far)}\sqrt{Q}-\rho_{B}\left( \frac{1+\sqrt{1-\hat{Q}}}{\hat{Q}} \right) ^n &\left\{ 2^{n+1}+n\left( 1-\sqrt{1-\hat{Q}} \right) ^n\sqrt{\hat{Q}}\right. \\
 &\left.\times \text{Beta}\left[ \frac{1}{2}-\frac{1}{2}\sqrt{1-\hat{Q}},-\frac{1}{2}-n,\frac{1}{2} \right] \right\}~.~\label{Eq.early time f(Q)}
 \end{aligned}
\end{equation}

In the above expressions, the ${}_2\text{F}_1$ and \text{Beta} are the Hypergeometric function and Beta function correspondingly, and the notation $\hat{Q}\equiv Q/Q_{m}$ in which $Q_m=2\alpha/[3\left( 1+\gamma \right) ^2]$. In addition, for the case of LQC, we have $ \omega_{\text{eff}}=\gamma=0$, and the Lagrangian~\eqref{Eq.late time f(Q)} and~\eqref{Eq.early time f(Q)} gets reduced to the following forms respectively

For $-1\leqslant \sqrt{\alpha}\, t\leqslant1 $:
\begin{equation}
f_{\text{LQC}}\left( Q \right) =A^{(near-LQC)}\sqrt{Q}-\rho_{B}\left[ 1+\sqrt{1-\frac{3Q}{2\alpha}}+\sqrt{\frac{3Q}{2\alpha}} \arcsin \left( \sqrt{\frac{1}{2}-\frac{1}{2}\sqrt{1-\frac{3Q}{2\alpha}}}\right) \right]~,
\end{equation}

For $-\infty<\sqrt{\alpha}\, t \leqslant-1 $ and $ 1\leqslant \sqrt{\alpha}\, t< \infty$:
\begin{equation}
f_{\text{LQC}}\left( Q \right) =A^{(far-LQC)}\sqrt{Q}+\rho_{B}\left[ -1+\sqrt{1-\frac{3Q}{2\alpha}}+\sqrt{\frac{3Q}{2\alpha}} \arcsin \left( \sqrt{\frac{1}{2}-\frac{1}{2}\sqrt{1-\frac{3Q}{2\alpha}}}\right) \right]~.
\end{equation}

Next, we are interested in studying the evolution of the Hubble horizon of our model. In the scenario of~\eqref{Eq.scale factor}, the co-moving Hubble radius is given by
\begin{equation}
\left|r_H\left( t \right)\right| =\frac{1}{\left|aH\right|}=\frac{3\left( \gamma +1 \right)}{2a_B\alpha \left|t\right|}\left( 1+\alpha t^2 \right) ^{1-\frac{1}{3\left( \gamma +1 \right)}}~.
\end{equation}
We can distinguish it into three different cases

\begin{figure*}[htb]
\centering
\includegraphics[height=5cm,width=8cm]{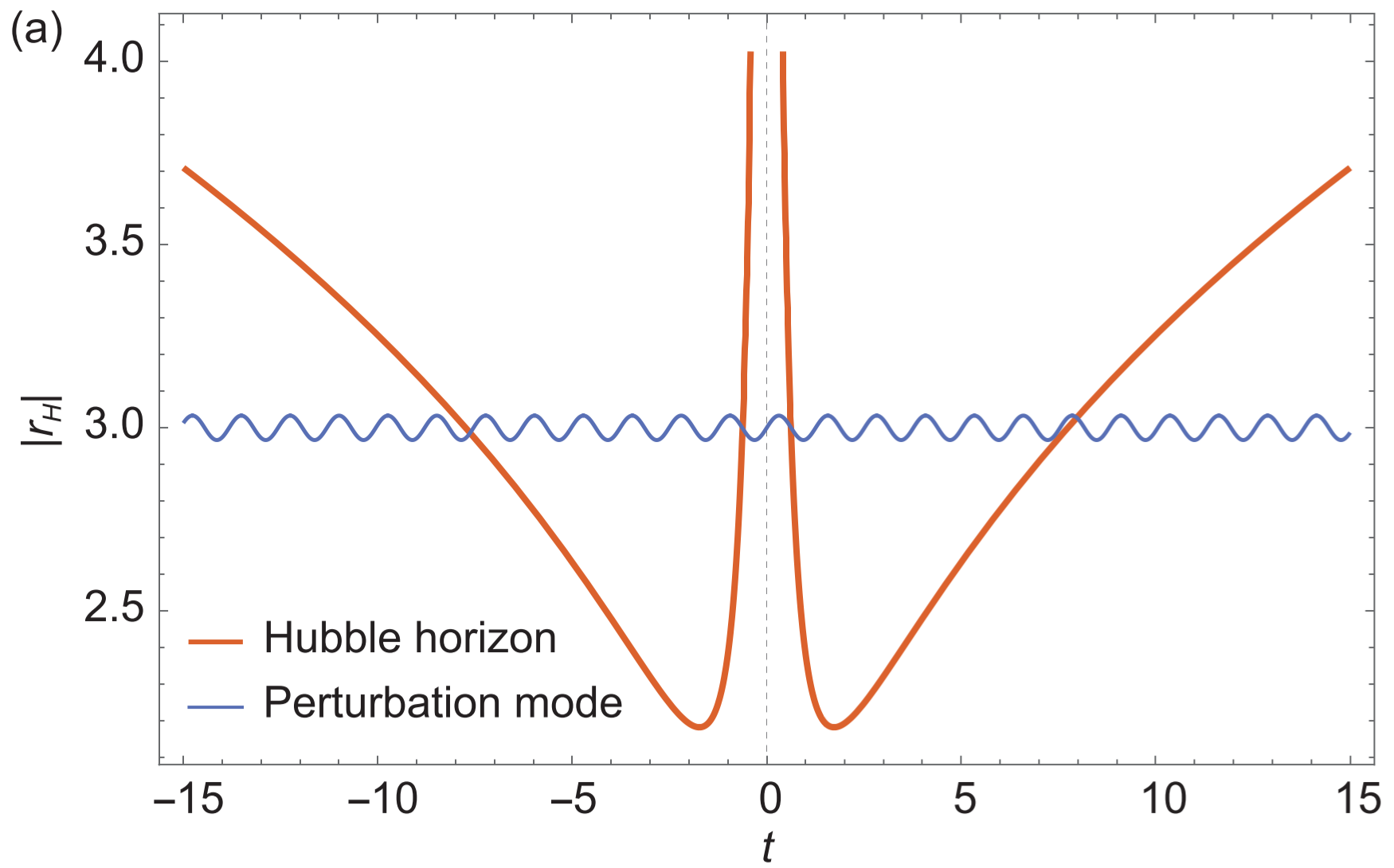}
\includegraphics[height=5cm,width=8cm]{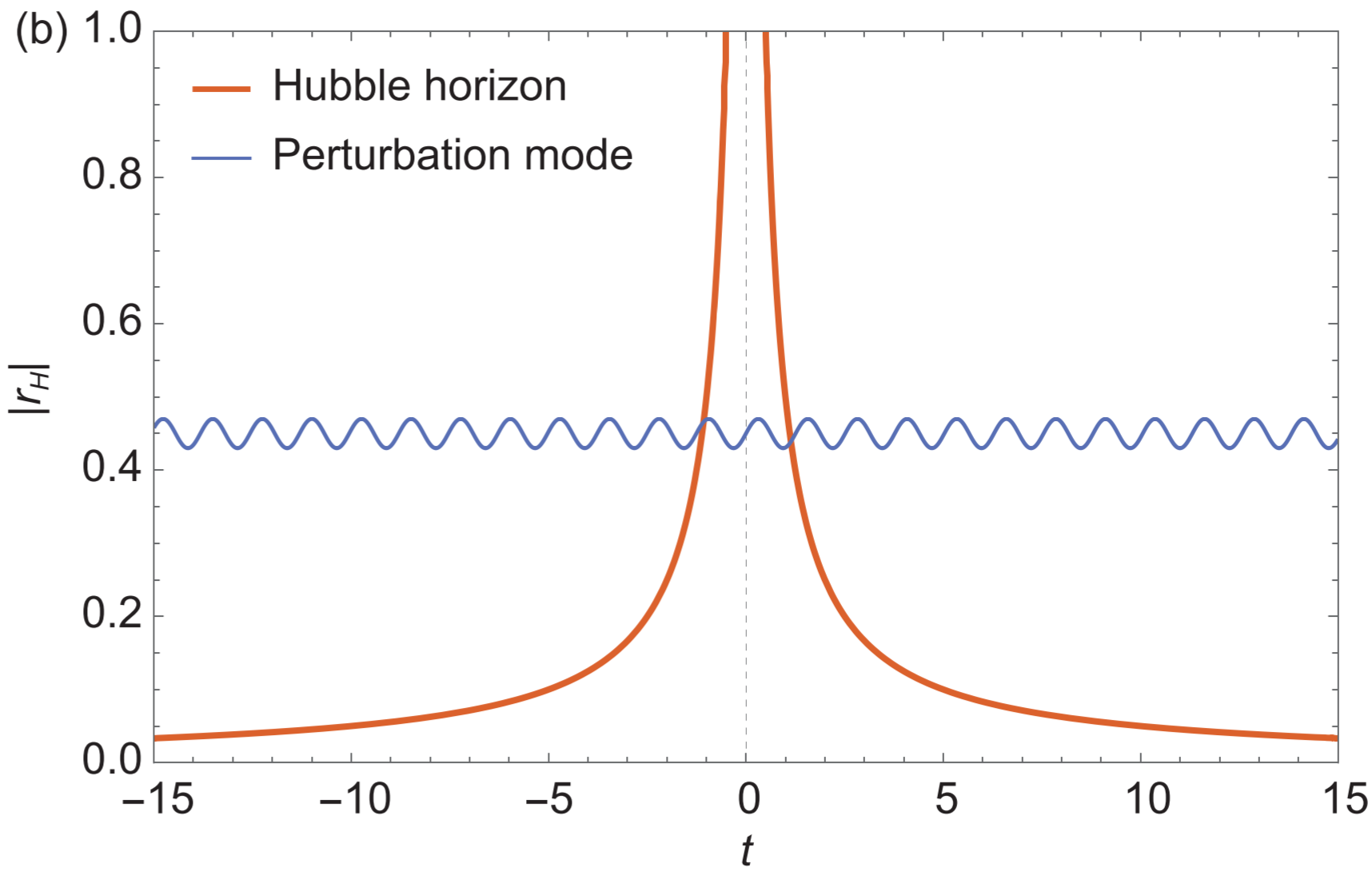}
\caption{The plot of the co-moving horizon of our model (red) and the fluctuation mode that cross the horizon before and after the bounce (blue). We choose the parameters $\alpha=1, \gamma=0, a_{B}=1$ (a), and $\alpha=1, \gamma=-2/3, a_{B}=1$ (b).} 
\label{Horizon}
\end{figure*}

For the case of $\gamma >-1/3$, as shown in the left panel of  FIG.~\ref{Horizon}., the primordial perturbation modes generate far away from the bounce in the deep contracting era. Along with the contraction of the Universe, the quantum fluctuations exit the Hubble radius because the horizon decreases fast near the bounce. After passing the bounce point, it will stay outside of the radius until re-enter the horizon at some point in the late-time expansion era.

For $-1<\gamma \leqslant -1/3$, as shown in the right panel of  FIG.~\ref{Horizon}., when $t>0$ the Hubble radius decreases monotonically over time and eventually goes to zero in the far future. Thus the primordial perturbation modes generate near the bounce area where $r_{H}\rightarrow\infty$, and it will stay outside of the horizon till today. As a result, the perturbation modes in the distant past lie in the super-Hubble regime, and thus the resolution of the horizon problem is problematic in this case. Thereby this situation should be excluded.

For $\gamma <-1$, the horizon evolves similarly to the case of $\gamma >-1/3$. However, the scale factor would permanently decrease after the bounce and equal to zero today, this case should also be ruled out.

Therefore in order to have a viable bounce scenario, here we consider the parameter $\gamma$ (present in Eq.~\eqref{Eq.scale factor}) to be restricted by $\gamma >-1/3$ in the subsequent calculations.

\section{ Tensor Perturbations in $f(Q)$ bounce}
\label{sec4}
\subsection{Perturbative action}

In this section, we study the cosmological linear tensorial perturbations generated from the $f(Q)$ bounce model. In order to do this, we work with the comoving time $\eta \equiv \int{dt/a(t)}$ and consider the tensorial perturbation around the background metric \eqref{Eq.FLRW}, namely:
\begin{equation}
ds^2=a^2(\eta)[-d\eta^2+(\delta _{ij}+h_{ij})dx^idx^j]~,~\label{Eq.FLRW2}   
\end{equation}
where $h_{ij}$ is traceless and divergent-free. Subsituting this perturbed metric \eqref{Eq.FLRW2} into the action \eqref{Eq.action} one can straightforwardly get the second order action as~\cite{BeltranJimenez:2019tme}
\begin{equation}
\mathcal{S}_{\text {tensor }}^{(2)}=\frac{1}{2} \sum_\lambda \int  a^2(\eta) f_Q\left[\left(h_{(\lambda)}^{\prime}\right)^2-k^2 h_{(\lambda)}^2\right]\mathrm{d}^3 k \mathrm{~d}\eta~.~\label{Eq.second order action}
\end{equation}
The $h_{(\lambda)}(\eta)$ are eigenmode functions of $h_{ij}$ in Fourier modes while $\lambda=1,2$ represent its two helicity modes, and $k$ is the comoving wavenumber of planewave perturbations. Moreover, the overprime $'$ denotes differentiation with respect to comoving time $\eta$. It can be observed that both the gravity polarization modes evolve in an equal footing. This indicates that $f(Q)$ cosmology does not introduce any parity-violating terms in the gravity wave action. By taking the variation of Eq.~\eqref{Eq.second order action}, one can obtain the equation of motion of ${h}_{(\lambda)}$:
\begin{equation}
h''_{\left( \lambda \right)} +\left( \frac{2a'}{a}+\frac{1}{f_Q}\frac{\partial f_Q}{\partial \eta} \right) h'_{\left( \lambda \right)} +k^2h_{\left( \lambda \right)} =0~.~\label{Eq.eom of h1}
\end{equation}
With respect to Mukhanov-Sasaki variable $v_{\left( \lambda \right)}\left( \eta \right)=z(\eta)h_{(\lambda)}$ with $z^2(\eta)=a^2(\eta)f_Q$, the above equation can be recast as
\begin{equation}
v ''_{\left( \lambda \right)}\left( \eta \right) +\left( k^2-\frac{1}{z}\frac{d^2z}{d\eta ^2} \right) v_{\left( \lambda \right)}\left( \eta \right) =0~.~\label{Eq.eom of h2}
\end{equation}

\subsection{Stability analysis}\label{sec-stability}
The present subsection is reserved for investigating the stability of the tensor perturbation, which is ensured if the kinetic term of the perturbed action in \eqref{Eq.eom of h1} remains positive during the entire cosmic era, i.e.,
\begin{equation}
\frac{\partial f\left( Q \right)}{\partial Q}>0 ~,~\text{for}~ -\infty <t<\infty~.
\end{equation}

Clearly, the stability condition can be checked with the form of $f(Q)$ that we have determined in Eq.~\eqref{Eq.late time f(Q)} and~\eqref{Eq.early time f(Q)}. Since the value of $Q$ is much smaller than $Q_m$ either when the Universe is close to bounce or far from bounce, for simplicity, we expand our action up to the first order for the $Q/Q_{m}$. 

For $-1\leqslant \sqrt{\alpha}\, t\leqslant1 $, i.e, near the bounce,
\begin{equation}
\begin{aligned}
f\left( Q \right) &\simeq  A^{(near)}\sqrt{Q}-2\rho _B+\frac{nQ\rho _B}{2Q_m}+\mathcal{O}\left(Q^2\right)~,\\
f_Q\left( Q \right) &\simeq  \frac{A^{(near)}}{2\sqrt{Q}}+\frac{n\rho _B}{2Q_m}+\mathcal{O}\left(Q\right)~.~\label{Eq.action near to bounce}
\end{aligned}
\end{equation}

Since the $Q_{m},\rho_{B}$ and $Q(t)$ are all positive factors, the requirement for $f_{Q}>0$ near the bounce seems to be equivalent to the condition $A^{(near)}>-n\rho _B\sqrt{Q}/Q_m$. Since $Q$ varies from $0$ to $Q_{m}$, thus for the case $n>0$, the constraint on $A$ immediately reduces to $A^{(near)}>0$. However, for $n\leq 0$, the condition becomes $A^{(near)}>|n|\rho _B/\sqrt{Q_m}$.

For $-\infty<\sqrt{\alpha}\, t \leqslant-1 $ and $ 1\leqslant \sqrt{\alpha}\, t< \infty$, i. e. far from the bounce,
\begin{equation}
\begin{aligned}
f\left( Q \right) &\simeq  A^{(far)}\sqrt{Q}-4^n\rho _B\left( \frac{Q_m}{Q} \right) ^n\left( \frac{2}{1+2n}\right)+\mathcal{O}\left(Q^{-n+1}\right)~, \\
f_Q\left( Q \right) &\simeq  \frac{A^{(far)}}{2\sqrt{Q}}+4^n\frac{2n\rho _B}{\left( 1+2n \right) Q}\left( \frac{Q_m}{Q} \right)^n+\mathcal{O}\left(Q^{-n}\right)~.~\label{Eq.action far from bounce}
\end{aligned}
\end{equation}

In order to avoid the tensor perturbation instability far from the bounce, the condition for the integration constant $A^{(far)}>-4^{n+1}n\rho _BQ^{-1/2}(Q_m/Q)^n/(1+2n)$ must be satisfied. Thus from the perspective of the stability condition of the tensor perturbation during $-\infty<\sqrt{\alpha}\, t \leqslant-1 $ and $ 1\leqslant \sqrt{\alpha}\, t< \infty$, we get three different cases depending on the value of $n$ as: for the cases $n>0$ or $n<-1/2$, we get $A^{(far)}>-\frac{4^{n+1}n\rho _B}{\left( 1+2n \right) \sqrt{Q_m}}$; however, for $-1/2<n<0$, the tensor perturbation is not stable for any value of $A^{(far)}$. We summarize our results in Table. \ref{tab:Range of model values}.

\begin{table}[htbp]
    \caption{Value Ranges of the model parameters from stability analysis.}
    \label{tab:Range of model values}
    \centering
    \begin{tabular}{cccc}
      \toprule
      \multirow{2}*{$A$ vs. $n$} & & $A^{(near)}$ & $A^{(far)}$\\ \cmidrule(lr){2-4}
                    & & Near the bounce & Far from the bounce\\ \midrule
      \rule{0pt}{16pt}$n< -1/2$ & & $> \frac{\left| n \right|\rho _B}{\sqrt{Q_m}}$ & $> -\frac{4^{n+1}n\rho _B}{\left( 1+2n \right) \sqrt{Q_m}}$ \\
      \rule{0pt}{16pt}$-1/2\leqslant n\leqslant0$ & & $> \frac{\left| n \right|\rho _B}{\sqrt{Q_m}}$ & Unstable\\
      \rule{0pt}{16pt}$n>0$ & & $>0$ & $>-\frac{4^{n+1}n\rho _B}{\left( 1+2n \right) \sqrt{Q_m}}$
      \\\bottomrule
    \end{tabular}
\end{table}

\subsection{Evolution of tensor perturbations and power spectrum}

\begin{figure*}[h]
\centering
\includegraphics[height=10cm,width=14cm]{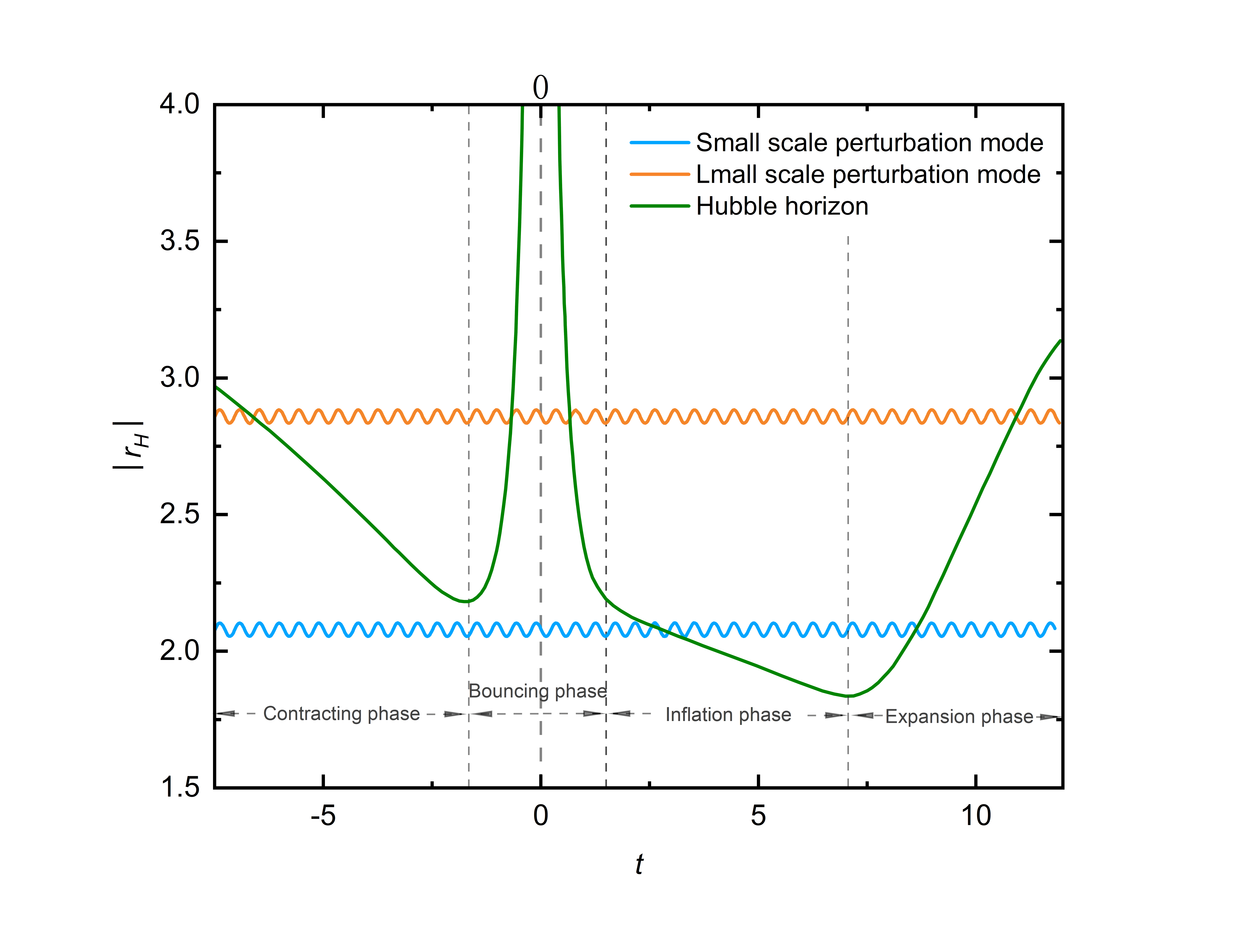}
\caption{The schematic diagram of the evolutionary behavior of the co-moving Hubble horizon in cosmic time $t$. The yellow wavy line represents perturbation mode with long wavelength, while the blue wavy line represents perturbation mode with short wavelength.} 
\label{sketch plot}
\end{figure*}

In this subsection, we study the evolution process of the tensor perturbations in our model. To be more precise, in order to solve the early Universe issues like flatness, horizon, and monopole problems as well, we consider the universe enters a period of inflation phase after the bounce, and this scenario is commonly known as bounce inflation~\cite{Piao:2003zm,Qiu:2015nha}. The inflationary phase may be realized by considering a scalar field (or inflaton) with non-zero potential in the current $f(Q)$ scenario over the normal matter fields. For instance, one may consider that the EoS parameter of the scalar field remains less than that of the normal matter field during cosmic evolution. As a result, the scalar field energy density remains suppressed in the contraction phase; however, after the bounce, the matter field energy gets diluted with the expansion of the Universe and the scalar field (or the inflaton) becomes the dominant component. Such inflaton field, along with the $f(Q)$ given in Eq.~\eqref{Eq.action far from bounce}, triggers an inflationary phase after the bounce in the Universe's evolution. We briefly sketch the evolution plot of the Universe in Fig.~\ref{sketch plot}. Having set the stage, we calculate how the tensor perturbations evolve during the whole process.  

\subsubsection{In contracting phase}

At the deep contracting phase when all the perturbation modes are initially submerged inside the Hubble radius, i.e when $k^2 \gg  z''/z$, the last term in the left-hand side of Eq.~\eqref{Eq.eom of h2} can be neglected. As a result, the tensor perturbation variable starts from the Bunch-Davies vacuum which corresponds to
\begin{equation}
v_{k}(\eta \rightarrow -\infty) = \frac{1}{\sqrt{2 k}} e^{-i k \eta}~.~\label{Eq.initial condition}
\end{equation}
Here we omit the subscript $(\lambda)$ for $v_{(\lambda)}$ as both the polarization modes have similar evolution. By using such Bunch-Davies vacuum as the initial condition for $v_{k}(\eta)$, we now try to solve the full Mukhanov-Sasaki Eq. \eqref{Eq.eom of h2}. For this purpose, we need to determine $z(\eta) = a(\eta)f_{Q}$ (and hence $z''/z$) in the contracting phase where the form of $f(Q)$ is given by~\eqref{Eq.action far from bounce} and $a\left(t\right)\propto t^{\frac{2}{3(\gamma +1)}}$. With such form of the scale factor along with $d\eta =a^{-1}(t)dt$, we get $\eta-\tilde{\eta}_{B-} \propto t^{\frac{1+3\gamma}{3\left( \gamma +1 \right)}}$, where $\tilde{\eta}_{B-}\equiv\eta_{B-}-[(1+3\gamma){\cal H}_{B-}/2]^{-1}$, and $\eta_{B-}$ and ${\cal H}_{B-}$ are the conformal time and conformal Hubble parameter at the transition point from contraction to bounce era. 
Such expressions of $f_{Q}$ and $a(\eta)$ lead to the following $z(\eta)$ in the contracting phase:
\begin{equation}
\begin{aligned}
z\left( \eta \right) &=a\left( \eta \right) \sqrt{f_Q}\\
&\simeq  {\cal M}\cdot \left|\eta-\tilde{\eta}_{B-} \right| ^s~,~\label{Eq.z}
\end{aligned}
\end{equation}
where the exponent $s=\frac{2\left( 2+n \right)}{1+3\gamma}+n+1$, and ${\cal M}$ is a complicated function, see the footnote \footnote{The coefficient $M$ takes the form of  ${\cal M} =\alpha ^{\frac{1+s}{3\left( 1+\gamma \right)}}\frac{3^{-s}\left( 1+\gamma \right) ^{2+n}}{2\left( 1+3\gamma \right) ^{-s}}\left( \frac{a_B}{1+\gamma} \right) ^{1+s}\sqrt{\frac{2^{-n}3^{1+n}n\rho _B}{1+2n}(Q_{m})^{n}}$. By choosing $\omega_{m}=\gamma=0$, it reduces to ${\cal M} =\frac{\alpha a_B^3}{18}\sqrt{\frac{2\rho _B}{Q_m}}$.}. Here we set the integral constant $A^{(far)}=0$ for simplicity, according to the stability condition in Table. \ref{tab:Range of model values}. 

Thanks to Eq.~\eqref{Eq.z}, the Mukhanov-Sasaki Eq.~\eqref{Eq.eom of h2} turns out to be
\begin{equation}
v_k''+\left[ k^2-\frac{s\left( s-1 \right)}{(\eta-\tilde{\eta}_{B-})^2} \right] v_k=0~,~\label{Eq.eom of h3}
\end{equation}
by solving which we get $v_k(\eta)$ in terms of Hankel functions as follows:
\begin{equation}
v_k\left( \eta \right) =\sqrt{\left|\eta-\tilde{\eta}_{B-}  \right|}\left[ c_1(k)\cdot H^{(1)}_{\nu_1}\left( k\left|\eta-\tilde{\eta}_{B-}  \right| \right) +c_2(k)\cdot H^{(1)}_{-\nu_1}\left( k\left| \eta-\tilde{\eta}_{B-}  \right| \right) \right],\quad \nu_1=\frac{1}{2}\left( 1-2s \right)~,~\label{Eq.solution of v}
\end{equation}
where $c_1(k)$, $c_2(k)$ are integration constants. From Fig.~\ref{sketch plot} we can see that, in the very far past where the fluctuation modes are deep inside the horizon, one gets $k^2\gg |z''/z|$, and $v_k$ becomes:
\begin{equation}
    v_k(\eta)\simeq \sqrt{\frac{2}{\pi k}}\left[c_1\cos(k|\eta-\tilde{\eta}_{B-}|)+c_2\sin(k|\eta-\tilde\eta_{B-}|)\right] ~,
\end{equation}
On the other hand, the horizon shrinks rapidly to a minimum value as the bounce point is approached. In this case, some perturbation modes with small $k$, i.e., $k^2\ll \left| z''/z \right|$, will exit the horizon. For this kind of superhorizon modes, $v_k$ becomes:
\begin{equation}
\begin{aligned}
v_k\left( \eta \right)\simeq \frac{c_1\left( k\left| \eta-\tilde{\eta}_{B-}\right| \right) ^{\frac{1}{2}+\nu _1}}{2^{\nu _1}\Gamma(1+\nu_1)\sqrt{k}}+\frac{c_2\left( k\left| \eta-\tilde{\eta}_{B-}\right| \right) ^{\frac{1}{2}-\nu_1}}{2^{-\nu _1}\Gamma(1-\nu_1)\sqrt{k}} ~.
\end{aligned}
\end{equation}

As a whole, tensor perturbation in the contraction phase is obtained as
\begin{equation}
\begin{aligned}
\text{Superhorizon:}&\quad h_k\left( \eta \right)  = \frac{v_k\left( \eta \right)}{z\left( \eta \right)} \simeq \frac{c_1\left( k\left| \eta-\tilde{\eta}_{B-}\right| \right) ^{2\nu _1}}{2^{\nu _1}\Gamma(1+\nu_1)k^{\nu_1}{\cal M}}+\frac{c_2}{2^{-\nu _1}\Gamma(1-\nu_1)k^{\nu_1}{\cal M}}~,
\\
\text{Subhorizon:}&\quad  h_k\left( \eta \right) \simeq \frac{\left| \eta-\tilde{\eta}_{B-} \right|^{\nu_1 -1/2}}{\cal M}\sqrt{\frac{2}{ \pi k}}\left[ c_1\cos(k|\eta-\tilde{\eta}_{B-}|)+c_2\sin(k|\eta-\tilde\eta_{B-}|) \right] ~.~\label{Eq.perturbation1}
\end{aligned}
\end{equation}
It may be observed that the subhorizon modes oscillate with a time-dependent amplitude. For the superhorizon modes, since $|\eta|$ is decreasing, for positive $\nu_1$ the second term in the expression of $h_k\left( \eta \right)$ decreases with the Universe's evolution and remains sub-dominant, hence the tensor perturbation gets frozen in the superhubble scale. However, in the LQC-like case where $n=-1$, or equivalently $\gamma=\omega_m=0$, $\nu_1 = -3/2$ turns out to be negative, hence the first term in the superhorizon mode is increasing and dominant. Moreover, matching the subhorizon solution  \eqref{Eq.perturbation1} with the Bunch-Davis vacuum (coming from Eq. \eqref{Eq.initial condition}) leads one to get the exact value of the coefficients:
\begin{equation}
    c_1=\frac{\sqrt{\pi}}{2}~,~~~c_2=-i\frac{\sqrt{\pi}}{2}~.
\end{equation}

\subsubsection{Through the bounce}

When the co-moving Hubble horizon $r_{H}$ passes through its minimum value, it will rapidly increase to infinity at the bounce point. During this process, all the perturbation modes lie within the subhorizon regime until cross the bouncing phase. Considering the $f(Q)$ (near the bounce) determined in Eq.~\eqref{Eq.action near to bounce} and the scale factor~\eqref{Eq.scale factor}, we obtain
\begin{equation}
z\left( \eta \right) =\sqrt{\frac{n\rho _B}{2Q_m}+\frac{\sqrt{3}\left( 1+\gamma \right) A}{4\sqrt{2}\alpha a_B\left( \eta -\eta _B \right)}}\left( 1+\alpha a_B^2(\eta-\eta_B ) ^2 \right) ^{\frac{1}{3\left( 1+\gamma \right)}}\simeq \sqrt{\frac{\sqrt{3}\left( 1+\gamma \right) A}{4\sqrt{2}\alpha a_B\left( \eta -\eta _B \right)}} ~,~\label{Eq.z2}
\end{equation}
(where we set $A^{(near)}=A$) and consequently, the Mukhanov-Sasaki Eq.~\eqref{Eq.eom of h2} becomes (in the leading order of $\eta \rightarrow \eta_B$)
\begin{equation}
v_k''+\left[ k^2-\frac{3}{4\left( \eta -\eta _B \right) ^2} \right] v_k=0~.~\label{Eq.eom of h4}
\end{equation}
From the above equation we can see that, in the bounce phase, what is to be compared with $k^2$ (representing the wavelengths of the fluctuation modes) is {\it not} horizon, but the {\it effective potential}. This is different from the contracting or expanding phase. Moreover, since $A>0$ in the function of $f(Q)$ from the analysis in the previous subsection, $f(Q)$ is dominated by $\sqrt{Q}\sim |H|$ rather than $H^2$ in GR, the effective potential is no longer constant as in \cite{Cai:2007zv, Qiu:2010ch}, but proportional to $(\eta-\eta_B)^2$. This means for $k|\eta-\eta_B|\ll 1$ is also a case to be considered, even if the fluctuation modes is inside the horizon.

One can solve Eq. \eqref{Eq.eom of h4} to get $v_k$ for large and small $k$ modes. Eq. \eqref{Eq.eom of h4} gives:
\begin{equation}
v_k\left( \eta \right) =\sqrt{\left|\eta-\eta_B  \right|}\left[ d_1(k)\cdot H^{(1)}_{1}\left( k\left|\eta-\eta_B  \right| \right) +d_2(k)\cdot H^{(1)}_{-1}\left( k\left| \eta-\eta_B \right| \right) \right]~,~\label{Eq.solution of v bounce}
\end{equation}
and for small $k$ modes ($k|\eta-\eta_B|\ll 1$),
\begin{equation}
 v_k \simeq \frac{d^{(s)}_1 (k|\eta-\eta_B|)^{\frac{3}{2}}}{2\Gamma(2)\sqrt{k}}+\frac{2d^{(s)}_2(k|\eta-\eta_B|)^{-\frac{1}{2}}}{\Gamma(0)\sqrt{k}}~,
\end{equation} 
for large $k$ modes ($k|\eta-\eta_B|\gg 1$), 
\begin{equation}
    v_k\simeq \sqrt{\frac{2}{\pi k}}[d^{(l)}_1(k)\cos(k|\eta-\eta_B|)+d^{(l)}_2(k)\sin(k|\eta-\eta_B|)]~.
\end{equation}

Here we use the superscript ``$s$'' and ``$l$'' to distinguish the coefficients associated with small $k$ and large $k$, respectively. The corresponding curvature perturbation in the bouncing phase can be calculated as
\begin{equation}
\begin{aligned}
\text{Small $k$:}&\quad  h_k\left( \eta \right) \simeq \left(\frac{32}{3}\right)^\frac{1}{4}\sqrt{\frac{ \alpha a_B}{\left( 1+\gamma \right) A }}\left[\frac{d^{(s)}_1|\eta-\eta_B|^2}{2\Gamma(2)}k+\frac{2d^{(s)}_2}{\Gamma(0)k}\right]~,
\\
\text{Large $k$:}&\quad  h_k\left( \eta \right) \simeq \left(\frac{128}{3}\right)^\frac{1}{4}\sqrt{\frac{ \alpha a_B \left| \eta -\eta _B \right|}{\left( 1+\gamma \right) A k\pi}}[d^{(l)}_1(k)\cos(k|\eta-\eta_B|)+d^{(l)}_2(k)\sin(k|\eta-\eta_B|)]~.~\label{Eq.perturbation2 k}
\end{aligned}
\end{equation}

The solutions for the bouncing phase should be matched to that during the contracting phase. For this purpose, we will use the Hwang-Vishniac (Deruelle-Mukhanov) matching condition \cite{Hwang:1991an, Deruelle:1995kd}, which requires that the $h_k$ and its derivative $h_k'$ should be continuous at the transition point of contraction-to-bounce phase, namely $t_{B-}=-1/\sqrt{\alpha}$ in the present model. Utilizing the solutions~\eqref{Eq.perturbation1} and~\eqref{Eq.perturbation2 k} we can determine the coefficients as follows\\
i) For small $k$:
\begin{equation}
\begin{aligned}
d_{1}^{\left( s \right)}\simeq &\left(\frac{3}{2}\right)^{1/4}c_1\left(\frac{k}{|{\cal H}_{B-}|}\right)^{-1+\nu_1}\left(\frac{2}{|{\cal H}_{B-}|}\right)^{\nu_1}\Gamma(1-\nu_1)\sin(\pi\nu_1)\sqrt{\frac{A(1+\gamma)a_B}{{\cal M}^2\pi^2(\Delta\eta)^2\alpha}} ~,
\\
d_{2}^{\left( s \right)}\simeq &2^{\nu_1}\left(\frac{3}{2^{17}}\right)^{1/4}c_1k^{1+\nu_1}|{\cal H}_{B-}|^{-2\nu_1}({\cal H}_{B-}\Delta\eta-4)\Gamma(-\nu_1)\Gamma(0)\sin(\pi\nu_1)\sqrt{\frac{A(1+\gamma)a_B}{{\cal M}^2\pi^2\alpha}}~,~(\nu_1<0)
\\
\simeq & 2^{\nu_1}\left(\frac{3}{2^{9}}\right)^{1/4}c_2k^{1-\nu_1}\Gamma(\nu_1)\Gamma(0)\sin(\pi\nu_1)\sqrt{\frac{A(1+\gamma)a_B}{{\cal M}^2\pi^2\alpha}}~.~(\nu_1>0)
\end{aligned}
\end{equation}
ii) For large $k$: 
\begin{equation}
\begin{aligned}
d_{1}^{\left( l \right)}\simeq & \left(\frac{3}{8}\right)^{1/4}\left(\frac{2}{|{\cal H}_{B-}|}\right)^{-1/2+\nu_1}\sqrt{\frac{A(1+\gamma)a_B}{\alpha\Delta\eta{\cal M}^2}}\left(c_1\cos{\left[\frac{k}{2}\left(\Delta\eta-\frac{4}{{\cal H}_{B-}}\right)\right]}+c_2\sin{\left[\frac{k}{2}\left(\Delta\eta-\frac{4}{{\cal H}_{B-}}\right)\right]}\right)~,
\\
d_{2}^{\left( l \right)}\simeq &\left(\frac{3}{8}\right)^{1/4}\left(\frac{2}{|{\cal H}_{B-}|}\right)^{-1/2+\nu_1}\sqrt{\frac{A(1+\gamma)a_B}{\alpha\Delta\eta{\cal M}^2}}\left(c_2\cos{\left[\frac{k}{2}\left(\Delta\eta-\frac{4}{{\cal H}_{B-}}\right)\right]}-c_1\sin{\left[\frac{k}{2}\left(\Delta\eta-\frac{4}{{\cal H}_{B-}}\right)\right]}\right) ~,
\end{aligned}
\end{equation}
where $\Delta\eta\equiv\eta_{B-}-\eta_{B+}\simeq 2(\eta_{B-}-\eta_B)$ is the duration of bounce phase in conformal time.

\subsubsection{In inflation phase}

As mentioned earlier that after the bounce i.e. during $\sqrt{\alpha}t \geqslant 1$, the inflaton field becomes the dominating energy component of the Universe, which, along with the $f(Q)$ given in Eq.~\eqref{Eq.action far from bounce}, triggers the inflation process for a suitable choice of inflaton potential. During the inflation era, the Universe expands almost exponentially, while the Hubble parameter $H$ is nearly constant. Furthermore, the transition point from bounce-to-inflation locates at $t_{B+}=1/\sqrt{\alpha}$, because the Hubble horizon $1/H(t)$ given by Eq.~\eqref{Eq.scale factor} has a minimum value when $t_{B+}=1/\sqrt{\alpha}$, which makes smooth conjunction between the bounce and the inflation. 

For convenience, we assume that the scale factor during inflation takes the following form
\begin{equation}
a_{\text{inf}}\left( t \right) \simeq  \kappa \left( 1+\epsilon\cdot\beta\cdot t \right) ^{1/\epsilon}~,~0<\epsilon \ll 1~,~\label{Eq.infscalefactor}
\end{equation}
where $\beta$ and $\kappa$ are the parameters used to adjust the speed and the amplitude of the inflation, and the slow-roll parameter $\epsilon$ is defined by $\epsilon=-\dot H/H^2$. In order to satisfy the continuity from bounce-to-inflation, i.e. the matching of the Hubble parameters and scale factors given by Eq.~\eqref{Eq.infscalefactor} and Eq.~\eqref{Eq.scale factor} at $t_{B+}$, there should be intrinsic relationships between the parameters as the following:
\begin{eqnarray}
\beta\simeq\frac{\sqrt{\alpha}}{3\left( \gamma +1 \right)}~,~~~~~~\kappa =2^{\frac{1}{3\left( 1+\gamma \right)}}a_B\left(1 + \frac{\epsilon}{3(1+\gamma)}\right)^{-1/\epsilon}~~,\nonumber
\end{eqnarray}
where we consider $1 + \epsilon \approx 1$. Furthermore, the conformal time during the inflation is given by
\begin{equation}
\eta -\tilde{\eta}_{B+}=\frac{(1+\epsilon\beta t)^{\frac{\epsilon-1}{\epsilon}}}{\kappa\beta(\epsilon-1)}
\end{equation}
where $\tilde{\eta}_{B+}\equiv \eta_{B+}-[(\epsilon-1){\cal H}_{B+}]^{-1}$, and $\eta_{B+}$ and ${\cal H}_{B+}$ are the conformal time and conformal Hubble parameter at the transition point from bounce to inflation era. Having this, we can determine the inflationary scale factor~\eqref{Eq.infscalefactor} in terms of the conformal time: 
\begin{equation}
a_{\text{inf}}\left( \eta \right) \simeq \kappa[\kappa \beta (\epsilon -1)(\eta -\tilde{\eta}_{B+})]^{\frac{1}{\epsilon-1}}~,
\end{equation}
The above expression of $a(\eta)$ and the $f(Q)$ (given by Eq.~\eqref{Eq.action far from bounce}) immediately leads to 
\begin{equation}
z\left( \eta \right) \simeq {\cal N}\cdot | \eta -\tilde{\eta}_{B+} | ^{\frac{\epsilon (n+1)+1}{\epsilon -1}} 
\end{equation}
while ${\cal N}=\kappa ^{\frac{\epsilon (n+2)}{\epsilon-1}}\beta^{\frac{n+2}{\epsilon-1}} \sqrt{\frac{2^n n\left(Q_m\right)^n \rho_B}{3^{n+1}(1+2 n)}}$. For $\omega_m=\gamma=0$, it reduces to ${\cal N}=3\sqrt{\rho_B/(2\alpha Q_m)}$. The Mukhanov-Sasaki Eq.~\eqref{Eq.eom of h2} in inflation phase becomes 
\begin{equation}
v_k''+\left[ k^2-\frac{2+3\left( n+2 \right) \epsilon}{(\eta -\tilde{\eta}_{B+}) ^2} \right] v_k=0~,~\label{Eq.eom of h5}
\end{equation}
by solving which we get
\begin{equation}
v_k\left( \eta \right) =\sqrt{\left| \eta -\tilde{\eta}_{B+}  \right|}\left[ f_1 H^{(1)}_{\nu_2}\left( k\left| \eta -\tilde{\eta}_{B+}  \right| \right) +f_2 H^{(1)}_{-\nu_2}\left( k\left| \eta -\tilde{\eta}_{B+}  \right| \right) \right],\quad \nu_2 \simeq \frac{3}{2}+\left( n+2 \right) \epsilon ~.  ~\label{Eq.solution of v inf}
\end{equation}
Note that the above solution of $v_k(\eta)$ during the inflation is of the same form compared to that of in the contraction phase --- this is expected, because in both phases, the term $z''/z$ turns out to be proportional to $1/(\eta -\tilde{\eta}_{B+})^2$. 

What we are observing are the superhorizon modes at the end of the inflation phase, and they come in two different ways. As shown in Fig.~\ref{sketch plot}, there are two  branches of this solution: The first branch is the small $k$ modes, which have already exited
the horizon at the contracting phase. The second branch comes from the large $k$ modes, which are always subhorizon modes until they exit the Hubble horizon during the inflation phase. The solution for the first branch is
\begin{equation}
\begin{aligned}
v_k\left( \eta \right)\simeq &\frac{f^{(s)}_1\left( k\left| \eta-\tilde{\eta}_{B+}\right| \right) ^{\frac{1}{2}+\nu _2}}{2^{\nu _2}\Gamma(1+\nu_2)\sqrt{k}}+\frac{f^{(s)}_2\left( k\left| \eta-\tilde{\eta}_{B+}\right| \right) ^{\frac{1}{2}-\nu_2}}{2^{-\nu _2}\Gamma(1-\nu_2)\sqrt{k}} ~,
\\
h_k\left( \eta \right) \simeq &\frac{f^{(s)}_1\left( k\left| \eta-\tilde{\eta}_{B-}\right| \right) ^{2\nu _2}}{2^{\nu _2}\Gamma(1+\nu_2)k^{\nu_2}{\cal N}}+\frac{f^{(s)}_2}{2^{-\nu _2}\Gamma(1-\nu_2)k^{\nu_2}{\cal N}}~.
~\label{Eq.perturbation3 Smallk}
\end{aligned}
\end{equation}
The subhorizon modes solution for the second branch is
\begin{equation}
\begin{aligned}
v_k\left( \eta \right)\simeq &\sqrt{\frac{2}{\pi k}}\left[ f^{(l)}_1\cos(k|\eta-\tilde\eta_{B+}|)+f^{(l)}_2\sin(k|\eta-\tilde\eta_{B+}|) \right]~,
\\
h_k \left( \eta\right) \simeq & \frac{|\eta -\tilde{\eta}_{B+}|}{\cal N}\sqrt{\frac{2}{\pi k}}\left[ f^{(l)}_1\cos(k|\eta-\tilde\eta_{B+}|)+f^{(l)}_2\sin(k|\eta-\tilde\eta_{B+}|) \right] ~,~\label{Eq.perturbation3 Largek}
\end{aligned}
\end{equation}
The coefficients $f_1$, $f_2$ can be determined by the same method in the previous discussion. For the first branch, the solution~\eqref{Eq.perturbation3 Smallk} should be matched with the small $k$ modes in bouncing phase~\eqref{Eq.perturbation2 k}, by which we get:
\begin{equation}
\begin{aligned}
f_{1}^{\left( s \right)}\simeq &-\left(\frac{1}{6}\right)^{1/4}d^{(s)}_1\left(\frac{1}{2{\cal H}_{B+}(1-\epsilon)}\right)^{\frac{1}{2}-\nu_2}\left(\frac{k}{{\cal H}_{B+}(1-\epsilon)}\right)^{\frac{1}{2}-\nu_2}\Gamma(\nu_2)\sqrt{\frac{\alpha k (\Delta\eta)^2{\cal N}^2}{A(1+\gamma)a_B)}}  ~,
\\
\simeq & 2^{\nu_1+\nu_2-1}c_1\frac{{\cal N}}{{\cal M}\pi}|{\cal H}_{B-}|^{1-2\nu_1}k^{-\frac{1}{2}+\nu_1}\left(\frac{1}{{\cal H}_{B+}(1-\epsilon)}\right)^{\frac{1}{2}-\nu_2}\left(\frac{k}{{\cal H}_{B+}(1-\epsilon)}\right)^{\frac{1}{2}-\nu_2}\Gamma(1-\nu_1)\Gamma(\nu_2)\sin(\pi\nu_1)~,\\
f_{2}^{\left( s \right)}\simeq &-\left(\frac{1}{3\times 2^9}\right)^{1/4}(-k\Delta\eta)^{-\frac{3}{2}}\left(\frac{2}{{\cal H}_{B+}(1-\epsilon)}\right)^{\frac{1}{2}-\nu_2}\left(\frac{k}{{\cal H}_{B+}(1-\epsilon)}\right)^{\frac{1}{2}+\nu_2}\Gamma(-\nu_2)\sqrt{\frac{\alpha\Delta\eta {\cal N}^2}{A(1+\gamma)a_B}}\\
\times & \left(d^{(s)}_1(k\Delta\eta)^2[2+\nu_2(1-\epsilon){\cal H}_{B+}\Delta\eta]+16\frac{d^{(s)}_2}{\Gamma(0)}[\nu_2(1-\epsilon){\cal H}_{B+}\Delta\eta]\right)~,
\\
\simeq & 
2^{\nu_1-\nu_2-1}c_1\frac{{\cal N}}{{\cal M}\pi}k^{\nu_2-\nu_1}\left(\frac{k}{|{\cal H}_{B-}|}\right)^{2\nu_1}\Gamma(1-\nu_1)\Gamma(-\nu_2)\sin(\pi\nu_1)\left(\frac{{\cal H}_{B-}}{{\cal H}_{B+}(1-\epsilon)}+\frac{2\nu_2}{\nu_1}\right)~,~(\nu_1<0)
\\
\simeq &
2^{\nu_1-\nu_2}c_2\frac{{\cal N}}{{\cal M}\pi}k^{\nu_2-\nu_1}\Gamma(1-\nu_2)\Gamma(\nu_1)\sin(\pi\nu_1)~,~(\nu_1>0)
~\label{Small.f1f2}
\end{aligned}
\end{equation}
while the second branch of solution~\eqref{Eq.perturbation3 Largek} should be matched with the large $k$ modes in bouncing phase~\eqref{Eq.perturbation2 k}, from which we have:
\begin{equation}
\begin{aligned}
f_{1}^{\left( l \right)}\simeq &\left(\frac{8}{3}\right)^{1/4}[{\cal H}_{B+}(1-\epsilon)]^{\nu_2-\frac{1}{2}}\sqrt{\frac{\alpha\Delta\eta{\cal N}^2}{A(1+\gamma)a_B}}\left(d^{(l)}_1\cos\left[\frac{k\Delta\eta}{2}+\frac{k}{{\cal H}_{B+}(1-\epsilon)}\right]-d^{(l)}_2\sin\left[\frac{k\Delta\eta}{2}+\frac{k}{{\cal H}_{B+}(1-\epsilon)}\right] \right) ~,\\
\simeq & \left(\frac{2}{|{\cal H}_{B-}|}\right)^{-1/2+\nu_1}\frac{\cal N}{\cal M}[{\cal H}_{B+}(1-\epsilon)]^{\nu_2-\frac{1}{2}}\left(c_1\cos\left[k\left(\frac{2}{{\cal H}_{B-}}+\frac{1}{{\cal H}_{B+}(1-\epsilon)}\right)\right]-c_2\sin\left[k\left(\frac{2}{{\cal H}_{B-}}+\frac{1}{{\cal H}_{B+}(1-\epsilon)}\right)\right] \right)~,\\
f_{2}^{\left( l \right)}\simeq &\left(\frac{8}{3}\right)^{1/4}[{\cal H}_{B+}(1-\epsilon)]^{\nu_2-\frac{1}{2}}\sqrt{\frac{\alpha\Delta\eta{\cal N}^2}{A(1+\gamma)a_B}}\left(d^{(l)}_2\cos\left[\frac{k\Delta\eta}{2}+\frac{k}{{\cal H}_{B+}(1-\epsilon)}\right]+d^{(l)}_1\sin\left[\frac{k\Delta\eta}{2}+\frac{k}{{\cal H}_{B+}(1-\epsilon)}\right] \right) ~, \\
\simeq & \left(\frac{2}{|{\cal H}_{B-}|}\right)^{-1/2+\nu_1}\frac{\cal N}{\cal M}[{\cal H}_{B+}(1-\epsilon)]^{\nu_2-\frac{1}{2}}\left(c_2\cos\left[k\left(\frac{2}{{\cal H}_{B-}}+\frac{1}{{\cal H}_{B+}(1-\epsilon)}\right)\right]+c_1\sin\left[k\left(\frac{2}{{\cal H}_{B-}}+\frac{1}{{\cal H}_{B+}(1-\epsilon)}\right)\right] \right)~.~\label{Large.f1f2}
\end{aligned}
\end{equation}
We notice that Eq.~\eqref{Eq.perturbation3 Smallk} contains two terms, of which the first term in the bracket depends on $\left| \eta \right|$ while the last term is a constant. Since the $\left| \eta \right|$ is decreasing during inflation, thus the first term would decay along the evolution of time for positive $\nu_2$. For instance, in the LQC-like case where $n=-1$, $\nu_2=(3/2+\epsilon)$ turns out to be positive, hence the last term becomes dominant.

Moreover, when the large-k modes exit the horizon, they will become superhorizon modes which will have the same form as in Eq.~\eqref{Eq.perturbation3 Smallk} (although with different coefficients), namely: 
\begin{equation}
\begin{aligned}
v_k\left( \eta \right)\simeq &\frac{f^{(l)}_1\left( k\left| \eta-\tilde{\eta}_{B+}\right| \right) ^{\frac{1}{2}+\nu _2}}{2^{\nu _2}\Gamma(1+\nu_2)\sqrt{k}}+\frac{f^{(l)}_2\left( k\left| \eta-\tilde{\eta}_{B+}\right| \right) ^{\frac{1}{2}-\nu_2}}{2^{-\nu _2}\Gamma(1-\nu_2)\sqrt{k}} ~,
\\
h_k\left( \eta \right) \simeq &\frac{f^{(l)}_1\left( k\left| \eta-\tilde{\eta}_{B-}\right| \right) ^{2\nu _2}}{2^{\nu _2}\Gamma(1+\nu_2)k^{\nu_2}{\cal N}}+\frac{f^{(l)}_2}{2^{-\nu _2}\Gamma(1-\nu_2)k^{\nu_2}{\cal N}}~.
~\label{Eq.perturbation3 Largek-2}
\end{aligned}
\end{equation}

\subsubsection{The tensor power spectrum}

With the above solutions in hand, we are ready to obtain the final tensor power spectrum. The tensor spectrum is defined as:
\begin{equation}
P_{T}(k)\equiv 2\frac{k^3}{\pi ^2}\left| h_k \right|^2=2\frac{k^3}{\pi ^2}\left| \frac{v_k}{z} \right|^2~,
\end{equation}
where the factor 2 corresponds to the two independent polarization modes of the gravitational waves. Note that for both the first and the second branch, we are observing the perturbation modes when they are in the superhubble regime in the inflationary phase. Using Eq.\eqref{Eq.perturbation3 Smallk} (or \eqref{Eq.perturbation3 Largek-2}), we have
\begin{equation}
P_{T}(k)\simeq 2^{1+2\nu _2}\frac{k^{3-2\nu _2}|f_2|^2}{\Gamma ^2(1-\nu_2){\cal N}^2\pi^2}  ~,~\label{power spectrum}
\end{equation}
where $\nu_2 \simeq 3/2+\left( n+2 \right) \epsilon$. So we get:
\\
i) For small $k$:
\begin{equation}
\begin{aligned}
 P_T^{(s)}= & \frac{2^{2\nu_1-3}|{\cal H}_{B-}|^{-4\nu_1}[{\cal H}_{B-}\nu_1+2{\cal H}_{B+}(1-\epsilon)\nu_2]^2\Gamma^2(-\nu_1)\sin^2(\pi \nu_1)}{{\cal H}_{B+}^2{\cal M}^2\pi^3(1-\epsilon)^2\nu_2^2}k^{3+2\nu_1}~,~(\nu_1<0)
 \\
 =& \frac{2^{2\nu_1-1}\Gamma^2(\nu_1)\sin^2(\pi\nu_1)}{{\cal M}^2\pi^3}k^{3-2\nu_1}~,~(\nu_1>0)~\label{Eq.Pt2}
\end{aligned}
\end{equation}

ii) For large $k$: 
\begin{equation}
 P_T^{(l)}=\frac{4^{\nu_1+\nu_2-1}|{\cal H}_{B-}|^{1-2\nu_1}[{\cal H}_{B+}(1-\epsilon)]^{2\nu_2-1}}{\pi{\cal M}^2 \Gamma^2(1-\nu_2)}k^{3-2\nu_2}.~\label{Eq.Pt2}
\end{equation}

Clearly, the power spectrum gets frozen in the superhorizon scale. Here we need to recall that both $f_1$ and $f_2$ depend on $k$, and they have different forms for small $k$ and large $k$ (see Eq.~\eqref{Small.f1f2} and Eq.~\eqref{Large.f1f2} respectively). This indicates that the tilt of the tensor perturbation ($n_T$) for small $k$ is different compared to that of large $k$. It is worth mentioning that the integral constant $A$ does not appear in the final power spectrum; this is due to the fact that $A$ appears as a global constant in the action~\eqref{Eq.action near to bounce} and does not change the structure of the action, thus it is not surprising that the final power spectrum does not contain $A$. In particular, $n_T$ is obtained as:
\begin{equation}
\begin{aligned}
\text{small $k$:}\quad n_T&\equiv \frac{d\text{ln} P_T}{d\text{ln}k}\simeq 3+2\nu_1\simeq 4-2s~,~\text{for}~  s\geqslant \frac{1}{2}~,
\\
n_T&\equiv \frac{d\text{ln} P_T}{d\text{ln}k}\simeq 3-2\nu_1\simeq 2+2s~,~\text{for}~ s\leqslant \frac{1}{2}~,
\\
\text{large $k$:}\quad n_T&=3-2\nu_2\simeq -2\left( 2+n \right) \epsilon~.
\end{aligned}
\label{N1}
\end{equation}

To derive the expression of $n_T$ for small $k$, we used $\nu_1 = \left(1 - 2s\right)/2$. An important point to be noted from the above expression is that for $n<-2$ (for which the perturbation remains stable, see Sec.[\ref{sec-stability}]), the tensor perturbation at large $k$ gets a positive tilt --- this interestingly hints an enhancement of gravitational waves' amplitude at large $k$ in the background of a bounce-inflation scenario. Moreover Eq.(\ref{N1}) also hints at a consistency of the present $f(Q)$ bounce model with the recent PTA data \cite{NANOGrav:2023gor, Antoniadis:2023rey, Reardon:2023gzh, Xu:2023wog}, as the tensor tilt for small $k$ may acquire positive value for suitable parameters \cite{Vagnozzi:2020gtf, Tahara:2020fmn, Kuroyanagi:2020sfw}. Such possibilities clearly indicate that primordial gravitational waves' spectrum at the present Universe (and its possibility to cross various sensitivity curves of GWs observatories) in the context of $f(Q)$ bounce-inflation is worthwhile to examine, and this is expected to study in the near future.  

For the special case related to LQC, after substituting all the required coefficients into the Eq.~\eqref{power spectrum} and by choosing $n=-1$ (or equivalently $\gamma=\omega_m=0$) and $\rho_B=4\alpha /3$, we acquire the following expression for the corresponding tensor power spectrum:

i) For small $k$: 
\begin{equation}
P_T^{(s)}=\frac{{\cal H}_{B-}^6[3{\cal H}_{B-}+2{\cal H}_{B+}((2\epsilon^2+\epsilon-3)]^2}{256{\cal H}_{B+}^2{\cal M}^2\pi^2(2\epsilon^2+\epsilon-3)^2}~,
~n_T^{(s)}= 0~.~\label{Eq.Pt1LQC}
\end{equation}

ii) For large $k$: 
\begin{equation}
 P_T^{(l)}=\frac{4^{\epsilon-1}|{\cal H}_{B-}|^4[{\cal H}_{B+}(1-\epsilon)]^{2+2\epsilon}}{\pi{\cal M}^2 \Gamma^2\left(-\frac{1}{2}-\epsilon\right)}k^{-2\epsilon}~,~n_T^{(l)}\simeq -2\epsilon~.~\label{Eq.Pt2LQC}
\end{equation}

We can see that regardless of whether the wavelength is small or large, the tensor spectrum index is nearly zero, implying a nearly scale-invariant power spectrum. This is consistent with the well-known results of matter bounce inflation models \cite{Finelli:2001sr, Cai:2008qb}. Another interesting feature is that the power spectrum~\eqref{Eq.Pt1LQC}~\eqref{Eq.Pt2LQC} is proportional to $\alpha$ (the $\cal M\sim \alpha $, $\cal{H}_{B+}$ and $\cal{H}_{B-}$ are proportional to $\sqrt{\alpha}$), hence those for different $k$ differs only in the amplitude of the power spectrum. From the above analysis, one can see that, $\alpha$ has triple identities: it can measure the duration of the bounce process of the Universe; indicate the peak value of Hubble parameter/nonmetricity scalar at the transition point of contraction-to-bounce/bounce-to-inflation; and is also related to the matter density at the bounce point, $\rho_B$. If we get the constraints on $\alpha$ from the detection of primordial gravitational waves, we can constrain these three quantities at the same time. We expect future observations such as AliCPT \cite{Li:2017lat, Li:2017drr} can help us in this direction.  \footnote{Actually, what AliCPT (and other PGW experiments) detects is the tensor/scalar ratio of the perturbations. This means that to get constraints on $\alpha$, we should know the scalar perturbation as well. However, the scalar perturbation is more difficult to deal with, due to the redundant degrees of freedom \cite{BeltranJimenez:2019tme, Hu:2022anq}. This issue is under investigation now.} 

We also find that when $\epsilon\rightarrow 0$, one can reach the approximation of $P^{(s)}_T\simeq 9\alpha^2/256\pi^2$, while $P^{(l)}_T\simeq \alpha^2/16\pi^2$, therefore $P^{(s)}_T=(9/16)P^{(l)}_T$. This means that the small $k$ mode of tensor perturbations will be suppressed at small $k$ (large scale) region. This is also a very common feature in usual matter bounce scenarios (for instance, see \cite{Xia:2014tda}), which can be tested with new CMB data in the near future.

\section{discussion and conclusions}
\label{sec5}

In this work, we propose a bounce inflation cosmological scenario in a $f(Q)$ modified gravity theory, where the Universe initially contracts with a bouncing-like behavior, and smoothly transits to an inflationary era which can be further connected to the standard Big-Bang cosmology. In such a scenario, the presence of the inflation (after the bounce) ensures the natural resolution of the flatness problem, while the horizon problem gets resolved due to the initial Bunch-Davies vacuum at the distant past where all the perturbation modes lie within the sub-Hubble regime. Using a suitable reconstruction technique, we obtain the form of $f(Q)$ corresponding to this scenario.

First, we presented a parametrized form of bounce solution preceded by an inflationary era, and reconstructed the $f(Q)$ action using the inverse-function method. We found that for the case of matter bounce where $\gamma=\omega_m=0$, our model behaves similarly to the Loop Quantum Cosmology scenario, where the Friedmann equation receives a correction term in proportional to $\rho_m^2$. 

Next, we analyze the cosmological perturbation of $f(Q)$ gravity in the case of coincident gauge. However, as we argued in~\cite{Hu:2022anq}, due to the large number of possible propagation modes, the number of the rest degrees of freedom are unknown except for the two tensor degrees of freedom. For this reason, we studied the tensor perturbation, starting from the contracting phase to the inflation phase.

We investigated the stability issue of the tensor perturbations and showed the range of viable parameters in order to have stable perturbations. Our results show that there exist suitable parameter ranges which ensure the ghost-free nature of the present model during the entire cosmic era. We then calculate the tensor modes of primordial perturbations generated from the Bunch-Davies vacuum in the contracting phase, both for the large and small $k$ modes. We present our result for general unspecified parameters as well as the specific case of matter bounce, namely $\gamma=\omega_m=0$. As a result, the tensor power spectrum in the current context shows the following characteristics --- (a) for the case of matter bounce, the spectrum is found to be frozen in the superhubble regime, presenting nearly scale-invariance in both large and small scales. This is consistent with the general result of matter bounce inflation scenario. Moreover, the amplitude of the tensor perturbation is proportional to the parameter $\alpha$, which represents the bouncing duration, the Hubble parameter and nonmetricity scalar at the transition points and the matter density at the bouncing point. For future detection of primordial gravitational waves such as AliCPT, CMB-S4 and LiteBIRD, this parameter can be furtherly constrained along with the scalar perturbations. (b) the tensor spectrum for the small modes $k$ as well as the large $k$ modes may also get a positive tilt depending on the model parameters, which can be consistent with recent stochastic GW data from recent PTA observations. This also hints at a rich possibility of GWs' enhancement and its detection in the future. 

As a side remark, let's briefly comment on the similarity of the matter bounce case in our model and the LQC model. There are many cosmological models studied in the context of LQC, including the matter bounce scenario in LQC~\cite{Wilson-Ewing:2012lmx}, the slow-roll inflation~\cite{Ashtekar:2011rm,Ashtekar:2009mm} and so on. People also found the relationship between LQC and $f(T)$ modified gravity theory~\cite{Haro:2014wha}. Here, our result may also help us to deepen our understanding of the connection between LQC and modified gravity; especially the symmetric teleparallel version of LQC. Although there may be a certain amount of accidents, it might hint that there can be some quantum gravity nature hidden inside the phenomenological models of these non-Riemannian gravity theories, such as torsion or nonmetricity theory.

Although we mainly focus on the tensor perturbations (gravitational waves) of the $f(Q)$ bounce scenario, there are also many extended topics that can be achieved in the follow-up study. It is definitely interesting and necessary to investigate the more complicated scalar perturbations of this scenario as well, which is undergoing now; It is also intriguing to check if our stability condition of Table.\ref{tab:Range of model values} still be satisfied within a relatively small scale of the Universe i.e. galactic scales, which can be done by analyzing the perturbative situation in the case of spherical symmetric space-time \cite{DAmbrosio:2021zpm, Lin:2021uqa, Calza:2022mwt, Wang:2021zaz}. We will present our progress and results on this subject in the near future.

\begin{acknowledgments}
We thank Taishi Katsuragawa for helpful discussions. Taotao Qiu and Kun Hu were supported by the National Key Research and Development Program of China (Grant No. 2021YFC2203100), and the National Natural Science Foundation of China (Grant No. 11875141). Affiliations $1$ and $2$ contributed
equally to this work as the first affiliation.

\end{acknowledgments}

\bibliographystyle{apsrev4-1}
\bibliography{References.bib}

\end{document}